\documentclass{aa}
\usepackage{graphicx}
\usepackage{amsmath}
\usepackage{hyperref}
\usepackage{txfonts}
\usepackage{natbib}
\bibpunct{(}{)}{;}{a}{}{,}

\begin{document}

\title{%
Metallicity and age gradients in round elliptical
galaxies\thanks{Based on observations collected with the 6m telescope
of the Special Astrophysical Observatory (SAO) of the Russian Academy
of Sciences (RAS) which is operated under the financial support of
Science Department of Russia (registration number 01-43).}
}

\titlerunning{%
Metallicity and age gradients in round elliptical galaxies
}

\author{%
M.~Baes\inst{1}
\and
O.~K.~Sil'chenko\inst{2,3}
\and
A.~V.~Moiseev\inst{4}
\and
E.~A.~Manakova\inst{2}
}

\authorrunning{%
M.~Baes et al.
}

\offprints{O.~K.~Sil'chenko, \email{olga@sai.msu.su}}

\institute{%
Sterrenkundig Observatorium, Universiteit Gent, Krijgslaan 281 S9, B-9000 Gent, Belgium
\and
Sternberg Astronomical Institute, University av. 13, Moscow 119992, Russia
\and
Isaac Newton Institute of Chile, Moscow Branch
\and
Special Astrophysical Observatory, Nizhnij Arkhyz, 369167 Russia
}

\date{Received Nov 16, 2006 / Accepted Jan 26, 2007}

\abstract
{}
{%
We probe the stellar population age and metallicity distributions in nearby
elliptical  galaxies over the largest extension to date.
}
{%
Long-slit spectroscopy is made by using the spectrograph SCORPIO of
the 6-m telescope of the Special Astrophysical Observatory of the
Russian Academy of Sciencies. The Lick indices H$\beta$, Mg\,b,
Fe\,5270, and Fe\,5335 are calculated along the slit up to radii of
1.3 to 3 $r_e$ in 4 galaxies and up to 0.5 $r_e$ in the fifth one.  The
comparison with evolutionary synthesis models of simple stellar
populations allows us to disentangle age and metallicity and to measure
both.
}
{%
We have found that the mean stellar age is constant along the radius
only in one galaxy out of 5. The other 4 galaxies demonstrate quite
different behaviour of the mean stellar age: the outer parts are older than
the centres in 3 cases and younger -- in one case. The metallicity
gradients cannot be approximated by a single power law over the full
radial extension in 4 galaxies of 5. The inner metallicity
gradients within 0.5~$r_e$ are all rather steep, steeper than --0.4
metallicity dex per radius dex, and are inconsistent with the origin
of the elliptical galaxies by a major merger.
}
{}

\maketitle

\section{Introduction}

Elliptical galaxies were described by \citet{1926ApJ....64..321H} as
smooth homogeneous stellar systems. However even early photometric
studies revealed the existence of colour gradients along the radius in
bright ellipticals \citep{1969AJ.....74..354T, 1974AJ.....79..835B},
which had been treated later as metallicity gradients because of
consistency between the metallicity slopes determined from
different-band colour gradients \citep{1978ApJ...223..707S,
1989Ap&SS.156..127P}. But there were also some hints, after comparing
colour gradients to absorption-line strength gradients, that there may
be also age variations along the radius that make a minor
contribution to the colour gradients \citep{1990MNRAS.245..217G}.

The presence of metallicity gradients along the radius in elliptical
galaxies is a crucial test for all models of galaxy formation. The
model of monolithic dissipative collapse of a protogalactic gaseous
cloud by \citet{1974MNRAS.166..585L} predicted the metallicity gradients
with the correct sign: during the violent star formation phase over the
whole galaxy, the gas enriched by fresh heavy elements fell into the
centre and became the material for subsequent star formation. The more
elaborated models of dissipative collapse by
\citet{1984ApJ...286..403C} gave the characteristic value for the
final stellar metallicity gradient in the largest elliptical galaxies
of $-0.5$ dex per radius dex, and the dependence of the gradient slope
on the galactic mass. Small elliptical galaxies had no
metallicity gradient in the models of \citet{1984ApJ...286..403C}, due to
the early occurrence of a galactic wind that stopped the galaxy
formation. The other main scenario in which elliptical galaxies are
formed is the framework of hierarchical merging. In simulations of
elliptical galaxies forming by major merging, metallicity gradients
are also predicted. In general, the gradients in these simulations are
shallower than those formed in the monolithic collapse scenario
\citep{1980MNRAS.191P...1W, 1996IAUS..171..191B, 1999ApJ...513..108B}.

Recently \citet{2004MNRAS.347..740K} has considered the evolution of
metallicity gradients in the framework  of a modern LCDM model of elliptical
galaxy formation.  The processes considered include monolithic
collapse at high redshift and subsequent minor and major mergers
followed in some cases by induced star-formation bursts. Monte Carlo
simulations have given 124 various histories of elliptical galaxy
evolution and reproduced the observed variety of metallicity
gradients. The extreme case of a galaxy formed by the pure monolithic
collapse and then evolving quietly corresponds to a very steep
metallicity gradient, about $\Delta$[Fe/H]$/\Delta \log r \approx
-1$. During the evolution, the metallicity gradients become shallower
when galaxies merge; no galaxy having experienced some major merger
has the metallicity gradient steeper than $-0.35$ dex per radius dex.
Among the galaxy ensemble simulated by \citet{2004MNRAS.347..740K} the
average metallicity gradient is $\Delta$[Fe/H]$/\Delta \log r
\approx -0.3$, with a scatter of $\pm 0.2$ dex -- quite consistent
with the observations of colour gradients in elliptical galaxies; no
correlation between gradient and galaxy mass is found.  These
simulations demonstrate that the knowledge of exact values of
metallicity gradients in elliptical galaxies is important for probing
their evolution. In particular, \citet{2004MNRAS.347..740K} suggests
that the metallicity gradient in any individual elliptical galaxy is
an indicator of its history and may reveal if major mergers were
significant during its evolution.

The current observational data provide mean metallicity gradients
$\Delta$[Fe/H]$/\Delta \log r$ typically in the range $-0.2$ to $-0.3$
\citep[see e.g.][]{1993MNRAS.265..553C, 1993MNRAS.262..650D,
1999ApJ...527..573K, 2003A&A...407..423M}, while observational
evidence of the predicted correlation between metallicity gradients
and galaxy mass or luminosity remains somewhat controversial
\citep{1988A&A...203..217V, 1990AJ....100.1091P, 2005MNRAS.361L...6F}.

Unfortunately, investigating the metallicity gradients in elliptical
galaxies is not straightforward, and these observational results
should be interpreted with caution. Indeed, most estimates are made
under the {\it a priori} assumption that the whole colour or
absorption-line gradient results from metallicity gradient.  For
example, we could refer to the paper by \citet{1999ApJ...527..573K}
who summarise a totality of data on metal-line index gradients in 80
elliptical galaxies over 20 independent studies and calculate
metallicity gradients by using evolutionary synthesis of simple
stellar populations (SSP) and by assuming that the stellar populations
at any radius in all 80 galaxies under consideration are 17 Gyr old
and no age gradient exists. But we know that age-metallicity
degeneracy concerns all the optical colours, as well as all the
metal-line indices. If one diminishes either metallicity under a constant
age or age under a constant metallicity, all the colours would be
bluer and all the metal lines would be weaker. So ages and metallicity
must be determined simultaneously by confronting for examples
metal-line indices to the Balmer lines or to the broad-band optical
colours, to make definite conclusions about the metallicity (and age)
gradients.  Secondly, due to the different scales one must be careful
when comparing observationally-determined metallicity gradients to
the results of simulation.  For example, the metallicity
gradients presented by \citet{2004MNRAS.347..740K} cover the region
out to $2\,r_{\text{e}}$, whereas most observational metallicity
gradients are derived over a relatively compact region, typically
within 0.5 to $1\,r_{\text{e}}$. Simple extrapolations of the measured
gradients to larger radii are very dangerous possibly leading to
erroneous conclusions.

This work presents a precise determination of metallicity and age gradients in
five elliptical galaxies made by using
Lick-index measurements and by breaking the age-metallicity degeneracy
with the two-index diagrams based on the evolutionary synthesis of old
SSPs. The other feature of this work is a significant extension of the index
profiles: for 4 of 5 galaxies we reach 1.3 to 3 effective radii,
whereas most works up to now have considered index gradients within one
effective radius. In Sect.~2 we present our sample and in Sect.~3
we discuss the observations, data reduction, and the analysis of the
spectra. In Sect.~4 we discuss the raw line-index profiles for the
galaxies in our sample and convert these indices to age and
metallicity gradients using SSP diagrams. These results are discussed
in Sect.~5. Section~6 offers a summary.

\section{Sample}

This work is part of the larger project involving complex
photometric and spectral studies of nearby round elliptical
galaxies. The ultimate aim is to probe the dynamical and stellar structures
of elliptical galaxies up to a few effective radii from the centre. Round
ellipticals were selected because of their probably spherical shape
so the simplest one. For this reason we suggest that
one longslit cross-section with an arbitrary orientation of the slit
would be enough to obtain a complete picture of radial variations for
all kinematical and stellar population properties.  The database
HYPERLEDA has been used to select all E0 galaxies with a visible
ellipticity less than 0.075 (at the 25th blue isophote) and with the
diameters, also at the 25th blue isophote, between 1 and 3 arcminutes.
The latter condition is imposed to provide sufficient spatial
resolution, on one hand, and to match the field of view of the
spectrograph SCORPIO, on the other . We must have blank sky
measurements for the external parts of the slit to subtract the sky
contribution properly. The northern part of the sample lists 18
galaxies. Now we present the first part of the results of the spectral
observations for 5 of them. The global characteristics of the
considered galaxies are given in Table~1.

\begin{table*}
\caption{Global parameters of the galaxies}
\centering
\begin{tabular}{lll|ccccc}
\hline\hline
& & & & & & & \\
               & unit         & source      & NGC\,2634 & NGC\,2810 & NGC\,3348 & NGC\,4339 & IC\,2179  \\
& & & & & & & \\ \hline
& & & & & & & \\
Type           &              & NED$^1$     & E1:& E & E0 & E0 & E1  \\
$V_r $         & km\,s$^{-1}$ & NED         & 2258 & 3572 & 2837& 1289 & 4336 \\
Distance       & Mpc          & LEDA$^2$    & 34.7 & 53.5 & 44.3 & 18.2(Virgo) & 64.6 \\
$D_{25}$       & arcmin       & LEDA        & 1.7& 1.7 & 2.0 & 2.4 & 1.1 \\
$D_{25}$       & kpc          & LEDA        & 17.2 & 26.5 & 25.8 & 12.7 & 20.7 \\
$B_T^0$        & mag          & RC3$^3$     & 12.71 & 13.07 & 11.84 & 12.31 & 13.21 \\
$M_B$          & mag          & LEDA        & --20.0 &  --20.7 & --21.7 & --19.3 & --20.9  \\
$(B-V)_T^0$    & mag          & RC3         & 0.90 & 0.95 & 0.95 & 0.90 & 0.93 \\
$(U-B)_T^0$    & mag          & RC3         & 0.48 & -- & 0.44 & 0.50 & 0.52 \\
$\sigma_0$     & km\,s$^{-1}$ & LEDA        & $182\pm4$ & $223\pm8$ & $ 238 \pm 10$ & $113 \pm 4$ & $146 \pm 52$ \\
$r_{\text{e}}$ & arcsec       & (see notes) & $18^4$ & $14^5$ & $22^4$ & $26^5$ & $15^5$ \\
Environment    &              & NOG$^6$     & group, & field? & group, & Virgo & group, \\
               &              &             & off-center &  &  centre  & cluster & near centre \\
& & & & & & & \\ \hline
\multicolumn{6}{l}{$^1$\rule{0pt}{11pt}\footnotesize
NASA/IPAC Extragalactic Database}\\
\multicolumn{6}{l}{$^2$\rule{0pt}{11pt}\footnotesize
Lyon-Meudon Extragalactic Database}\\
\multicolumn{6}{l}{$^3$\rule{0pt}{11pt}\footnotesize
Third Reference Catalogue of Bright Galaxies}\\
\multicolumn{6}{l}{$^4$\rule{0pt}{11pt}\footnotesize
\citet{2004AJ....127.1917T}}\\
\multicolumn{6}{l}{$^5$\rule{0pt}{11pt}\footnotesize
This work}\\
\multicolumn{6}{l}{$^6$\rule{0pt}{11pt}\footnotesize
\citet{2000ApJ...543..178G}}\\
\end{tabular}
\end{table*}

\section{Observations and data reduction}

Long-slit spectral observations were made with the multi-mode
focal reducer SCORPIO \citep{2005AstL...31..194A} installed at the
prime focus of the BTA~6-m telescope at the Special Astrophysical
Observatory. A description of the SCORPIO instrument can be found at
http://www.sao.ru/hq/moisav/scorpio/scorpio.html.  For our
observations with the SCORPIO we used the phase-volume
holographic grating (the grism) VPHG2300 providing the narrow
spectral range of 4800--5540~\AA; this wavelength region is rich in
stellar absorption lines and contains the Lick indices H$\beta$,
Mg\,b, Fe\,5270, and Fe\,5335 suitable for studying stellar
populations. Moreover, the emission lines
[OIII]\,$\lambda\lambda$4959, 5007 appeared to be strong in some cases;
they are used for the gas kinematics study and for estimating the
emission contamination level for the Lick index H$\beta$. The slit
width was one arcsec, and the spectral resolution about 2.2~\AA. The CCD
$2k \times 2k$ was used as a detector, and the scale across the
slit was $0.35\arcsec$ per pixel. The full log of observations is given
in the Table~2. Besides the galaxies, several G- and K-giant stars
from the list of \citet{1994ApJS...95..107W} were observed in the same
mode, and their spectra were used for the Lick system calibration and for
cross-correlation with the spectra of the galaxies.  The preliminary
data reduction was performed by means of IDL-based software
\citep{2005AstL...31..194A}.

\begin{table}
\caption{Long-slit spectroscopy of the galaxies}
\centering
\begin{tabular}{llccc}
\hline\hline \\
Date & Galaxy & Exposure & PA (slit) & Seeing \\
     &        & (min)    & (deg)     & (arcsec) \\ \\
\hline \\
6 Mar 05  & NGC\,3348 & 100 & 34 & 3 \\
7 Mar 05  & IC\,2179  & 220 & 38 & 3 \\
7 Mar 05  & NGC\,2810 & 120 & 63 & 3 \\
20 Apr 06 & NGC\,3348 & 120 & 293& 2 \\
21 Apr 06 & NGC\,4339 &  40 & 241& 2 \\
26 Apr 06 & NGC\,2634 & 280 & 45 & 4 \\ \\
\hline
\end{tabular}
\end{table}

During the  further reduction we applied logarithmic binning  along the slit
to provide a sufficient signal-to-noise ratio ($\text{S/N}>25-30$ per bin).
For every galaxy we calculated radial dependencies of the Lick
indices H$\beta$, Mg\,b, Fe\,5270, and Fe\,5335 that are suitable to
determine metallicity, age, and Mg/Fe ratio of old stellar populations
\citep{1994ApJS...94..687W}. To calibrate the SCORPIO index system
onto the standard Lick one, we observed 22 stars from the list of
\citet{1994ApJS...94..687W} during several observational runs with the
same grating and spectrograph setup; the comparison of the
instrumental and standard Lick indices for these stars from the work
of \citet{1994ApJS...94..687W} is shown in Fig.~1. We calculated
the linear regression formulae to transform our index measurements
into the Lick system. The rms scatters of points near the linear
dependencies derived are about 0.2~\AA\ for all 4 indices under
consideration, within the observational errors of
\citet{1994ApJS...94..687W}.

To correct the index measurements for the stellar velocity dispersion
which is usually substantial in the centres of early-type galaxies, we
smoothed the spectrum of the K3III standard star
\object{HD~167042} by a set of Gaussians of various widths, and the
derived dependencies of index corrections on $\sigma$ were
approximated by 4h-order polynomials and applied to the measured
index values before their calibrations into the Lick system. Four
galaxies of the five studied here have published central aperture
index data obtained in the original Lick system
\citep{1998ApJS..116....1T}. In Fig.~2 we compare our fully calibrated
central index measurements averaged between $-2.2\arcsec$ and
$+2.2\arcsec$ for every galaxy with the data of
\citet{1998ApJS..116....1T} taking in mind their aperture
of $2\arcsec\times 4\arcsec$. With 13 points of 16 within one sigma
from the lines of equality, and with other 3 points within two sigma,
the agreement seems to be good.

\begin{figure}
\resizebox{\hsize}{!}{\includegraphics{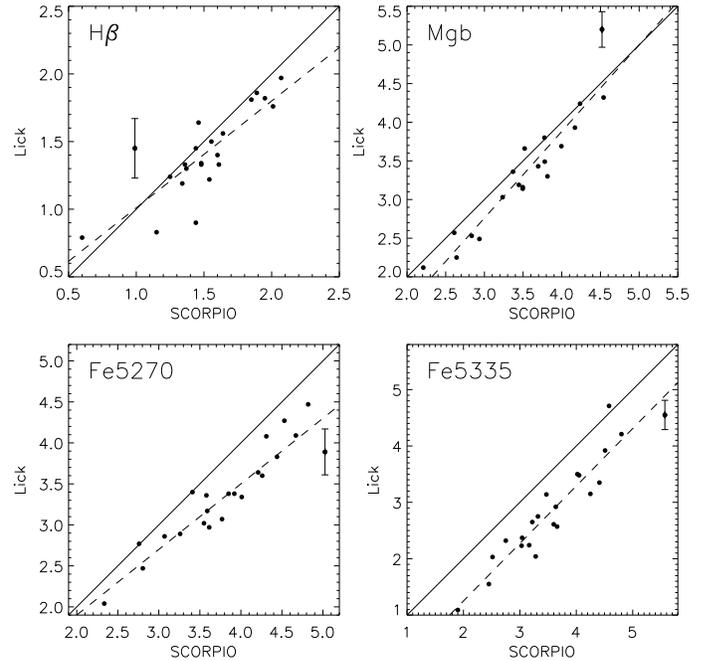}}
\caption{The correlation of the instrumental absorption-line indices
obtained with the spectrograph SCORPIO and the standard Lick indices
for 22 stars from the list of \citet{1994ApJS...94..687W}. The thin
straight lines are bisectrices of the quadrants,  and the dashed lines
linear regressions fitted to the dependencies. The mean index error
for the measurements of the stars by \citet{1994ApJS...94..687W} are
indicated for a single star in each plot; our errors are lower by one
order.} \label{indsys}
\end{figure}

\begin{figure}
\resizebox{\hsize}{!}{\includegraphics{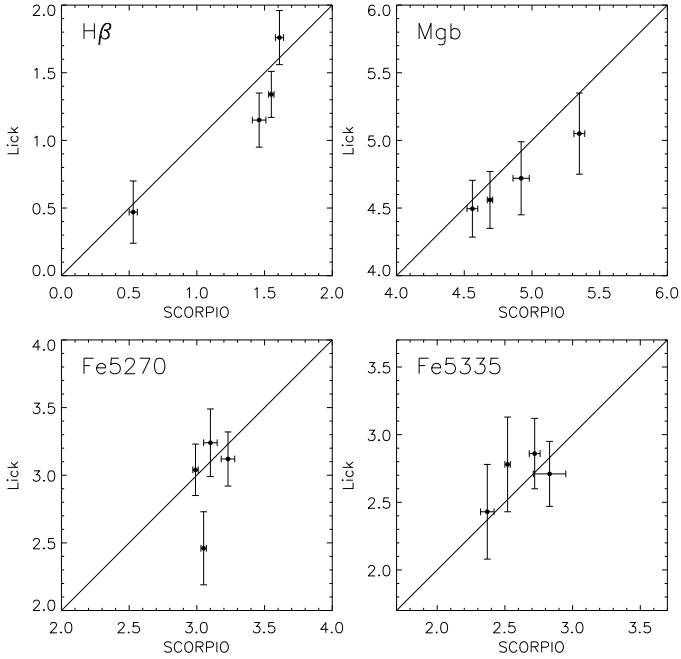}}
\caption{The comparison of the calibrated absorption-line indices
obtained with the spectrograph SCORPIO for the central parts
($1\arcsec \times 3\farcs5$) of four galaxies and the standard Lick
indices for them from the aperture spectroscopy of
\citet{1998ApJS..116....1T} with the aperture of $2\arcsec \times
4\arcsec$.} \label{indcomp}
\end{figure}

To evaluate the internal accuracy of our index measurements, we
made some simulations of the deepest galactic spectrum, the one for
NGC\,2634. The high S/N spectrum of the standard star
\object{HD~102328} was expanded along the slit with the
normalisation corresponding to the galaxy fluxes observed at different
distances from the centre, the Gaussian smoothing was applied to
imitate a radial velocity dispersion distribution, the sky level
and the noise observed were added, and after that our standard
reduction procedure was  applied to this artificial spectrum. By
measuring the indices, we assured that (i) difficulties related
to the sky subtraction affect only the index measurements in the
very external parts of the index profiles, beyond $2\,r_{\text{e}}$.
And (ii) the statistical accuracy of the indices estimated as the rms of
the point-to-point scatter is 0.02--0.04~\AA\ over the range 0 to
0.2\,$r_{\text{e}}$, 0.03--0.07~\AA\ over the range
0.2\,$r_{\text{e}}$ to 0.5\,$r_{\text{e}}$, 0.10--0.15~\AA\ over the
range 0.5\,$r_{\text{e}}$ to 1\,$r_{\text{e}}$, and 0.21--0.27~\AA\
over the range 1\,$r_{\text{e}}$ to 2\,$r_{\text{e}}$. As a visible
illustration of the high internal accuracy of our index measurements,
Fig.~3 compares two independent observations of
NGC\,3348, made in 2005 and in 2006 with the different slit
orientations; the agreement of the indices is within 0.2~\AA\ up to
$R=30\arcsec$, or $1.36\,r_{\text{e}}$.

\begin{figure}
\resizebox{\hsize}{!}{\includegraphics{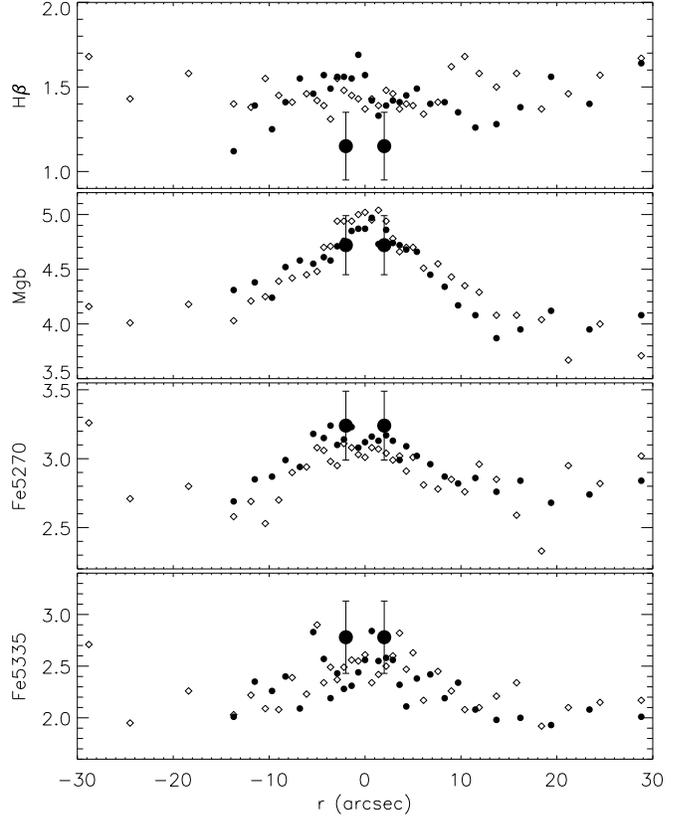}}
\caption{Two independent observations of NGC\,3348 in 2005 and in 2006
(small signs, filled and open) made with the orthogonal slit
orientations. The aperture central measurements by
\citet{1998ApJS..116....1T} are also shown by large filled circles
with error bars.}
\label{n3348fullprof}
\end{figure}

\section{The results}

\subsection{Line-index profiles}

Figures~4 to 8 show the measured index profiles for all 5 galaxies
under consideration. Two halves of the profiles are symmetrically
reflected with respect to the centre and plotted by different signs.
The first thing that is evident from Figs.~4 to 8 is that the index
radial variations cannot be fitted by a single linear law over the
full radius range, in opposition to well-established tradition. With
our high index accuracy, we clearly see breaks, or changes in slopes,
of the observed dependencies.  Below we give brief comments for every
galaxy.

\begin{figure}
\resizebox{\hsize}{!}{\includegraphics{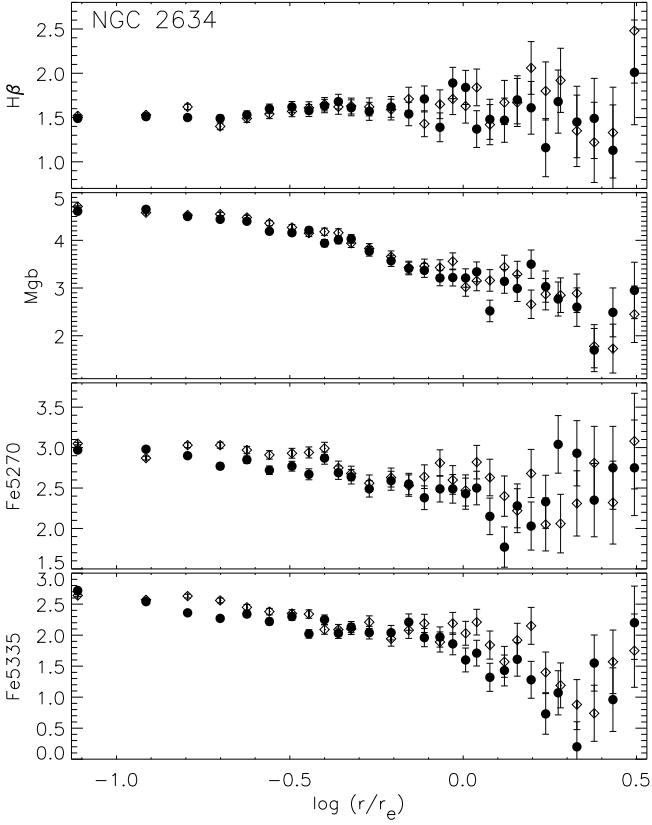}}
\caption{The index profiles of NGC\,2634.}
\label{n2634prof}
\end{figure}

{\bf NGC\,2634, Fig.~4.} This moderate-luminous elliptical galaxy
was observed by us with the longest exposure of them all, so the
observations are very deep. The index profiles are traced up to
$3\,r_{\text{e}}$, though we suspect that the sky subtraction uncertainty
may affect their shapes at $r>2\,r_{\text{e}}$. The emission
lines are absolutely absent in the spectrum, and the H$\beta$-index
profile looks  flat. The magnesium- and iron-index profiles
decline with radius. Their behaviour with respect to the logarithm
of the radius may be characterised as a sequence of linear pieces with
different slopes.  The points of slope breaks are close to
$0.4\,r_{\text{e}}$, or $\log (r/r_{\text{e}})=-0.4$, and, less
surely, to $0.8\,r_{\text{e}}$, or $\log (r/r_{\text{e}})=-0.1$.  This
break is seen in Mg\,b and Fe\,5335 profiles but not in the Fe\,5270
profile.  We take the value of $r_{\text{e}}$ for NGC\,2634 from
the work of \citet{2004AJ....127.1917T}, where its photometric structure
is described as a pure S\'ersic (non-core) one. The galaxy
is a member of large group dominated by a giant Sb-galaxy
\object{NGC\,2550}, but it is located rather far from the centre of the
group and has its own subgroup consisting of the dwarf
\object{NGC\,2634A} and of the late-type barred spiral
\object{NGC\,2633}.

\begin{figure}
\resizebox{\hsize}{!}{\includegraphics{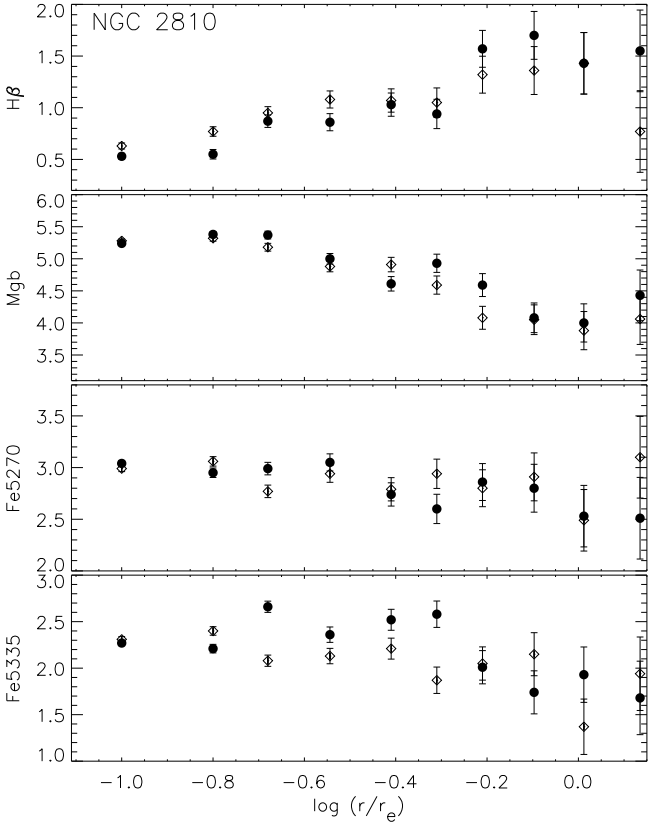}}
\caption{The index profiles of NGC\,2810.}
\label{n2810prof}
\end{figure}

{\bf NGC\,2810, Fig.~5.} The observations are shallower reaching only
$1.4\,r_{\text{e}}$. We determined the effective radius of the
galaxy from our SCORPIO data in  two ways: from the broad-band V-image
that always  precedes the spectral-mode exposure and from integration of
the flux obtained in the spectral mode across the slit. The surface
brightness profiles are well-fitted by a de Vaucouleurs' law \citep{1948AnnAp...11..247V},
and both approaches have consistently  given $r_{\text{e}} = 14\arcsec$. The
galaxy, though red and homogeneous photometrically,  revealed
strong emission lines in the spectrum, not only in the nucleus, but
also over the full radial extension.  Due to large difference between the
velocity dispersion of stars and ionized gas in the nucleus,
we succeeded in extracting a  pure emission spectrum and  estimated
the ratio of the emission-line fluxes: $\text{H}\beta /
\text{[\ion{O}{iii}]}\,\lambda 5007 = 0.85$. This ratio is within the
value interval found by \citet{2000AJ....119.1645T} for their sample
of elliptical galaxies, 0.33-1.25, so we used it to correct the
H$\beta$ indices measured in the spectrum of NGC\,2810 for the
emission through calculating $\text{EW}(\text{[\ion{O}{iii}]}\,\lambda
5007)$ over the full radial extension along the slit and multiplying it by 0.85.
However, Fig.~5 presents the initial, uncorrected H$\beta$ indices so
the H$\beta$ profile has a deep minimum in the centre of the galaxy.
The Mg\,b index, very high in the centre, falls steeply down to $\log
(r/r_{\text{e}}) \approx -0.1$ and stays roughly constant beyond this radius.
The iron index decline is rather shallow over the full
radius extension. The elliptical galaxy NGC\,2810 is attributed neither by
\citet{1993A&AS..100...47G} nor by \citet{2000ApJ...543..178G} to any
galactic groups, so in our Table~1 we classify it formally as a field
elliptical. But according to HYPERLEDA, a lot of small late-type
galaxies are seen not far from NGC\,2810; some of them are even
caught by the SCORPIO field of view.

\begin{figure}
\resizebox{\hsize}{!}{\includegraphics{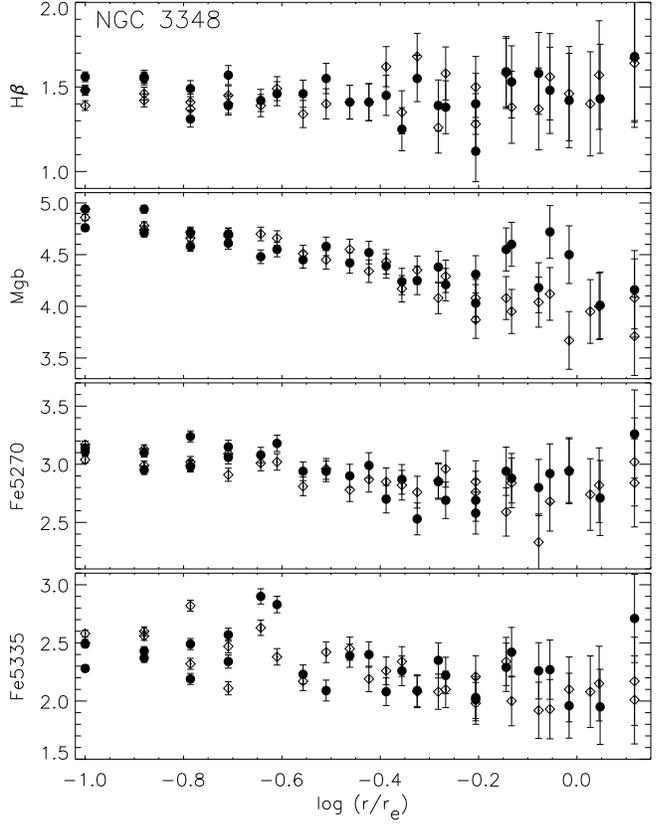}}
\caption{The index profiles of NGC\,3348.}
\label{n3348prof}
\end{figure}

{\bf NGC\,3348, Fig.~6.} This galaxy is the most luminous in our
sample, and it is the centre of its group. \citet{2004AJ....127.1917T}
describe it as a core elliptical galaxy, and it is  from this work
that we take the effective radius value of $22\arcsec$. The spectrum
demonstrates the weak emission line [\ion{O}{iii}]\,$\lambda$5007.
As we were not able to extract the pure emission line H$\beta$, we will
correct the index H$\beta$ below for the emission in the statististical
sense, by following the recommendation of \citet{2000AJ....119.1645T}:
$\Delta \text{H} \beta = 0.6 \text{EW}(\text{[\ion{O}{iii}]}\,\lambda
5007)$. Figure~6 presents an uncorrected H$\beta$ profile, as well as
magnesium- and iron-index profiles and combines the data for two
different exposures of the galaxy, that of March 2005 and that of
April 2006 made with the orthogonal orientations of the slit. The
metal-line profiles fall smoothly to $\log (r/r_{\text{e}}) \approx
-0.2$. Beyond this radius , the Mg\,b and Fe\,5335 profiles look flat,
and the Fe\,5270 profile even rises. The final points of the profiles
are around $r\approx 1.3\,r_{\text{e}}$. In April 2006,  the faint
galaxy at  $2\farcm 5$ to the east of NGC\,3348,
\object{MCG\,+12-10-079}, was also on the slit. It  appeared to be
the background early-type spiral galaxy, with the weak emission line
[\ion{O}{iii}]$\lambda$5007 in the spectrum indicating $v_r\approx
10200$ km s$^{-1}$. Its nuclear Lick indices, $\text{H}\beta = 1.38$,
$\text{Mg\,b}=3.32$, $\text{Fe}\,5270 = 3.03$, and $\text{Fe}\,5335 =
2.28$, are typical of  the bulge of the moderately luminous Sb galaxy.

\begin{figure}
\resizebox{\hsize}{!}{\includegraphics{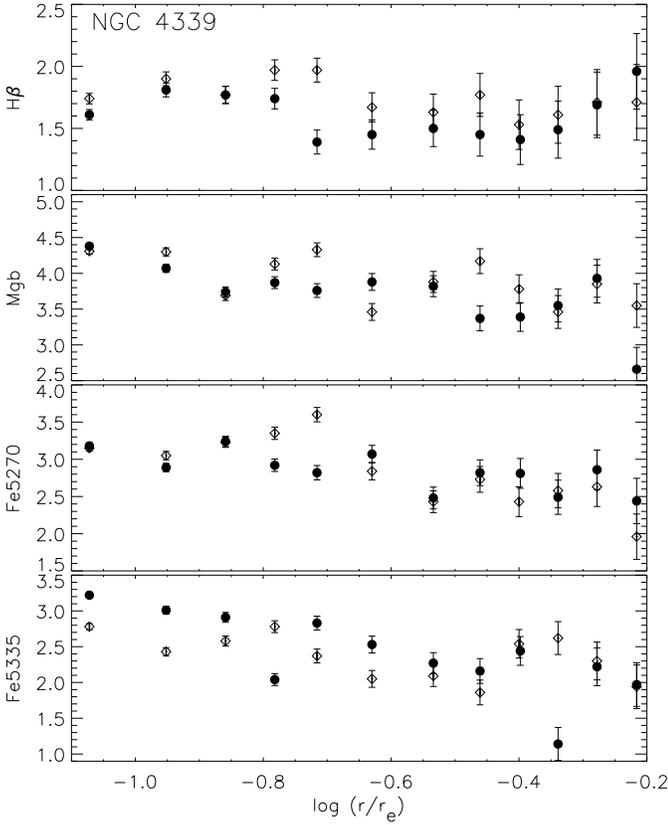}}
\caption{The index profiles of NGC\,4339.}
\label{n4339prof}
\end{figure}

{\bf NGC\,4339, Fig.~7.} This early-type galaxy is a Virgo member.
It has a rather low luminosity for an elliptical and is classified as
either E0 or  S0 by different authors. We analysed its surface
brightness profile obtained by tracing the flux integrated across the
slit and have seen that  the surface brightness profile beyond
$r\approx 20\arcsec$  is well-fitted by an exponential, so beyond
$r\approx 20\arcsec$ the stellar disk dominates. But within this
radius we see a de Vaucouleurs' bulge with $r_{\text{e}} =
26\arcsec$,  so this effective radius is used to parametrise the
index profiles in Fig.~7. \citet{1994A&AS..106..199C} give an
effective radius of $27\farcs7$ along the major and $25\farcs8$ along
the minor axes, so in agreement with our measurements.  Because of the
bad weather, the exposure of NGC\,4339 was  short, and we only
traced the index profiles for this galaxy to $0.5\,r_{\text{e}}$.
Nevertheless even these short profiles revealed  the
breaks of slopes, this time around $\log (r/r_{\text{e}}) \approx -0.7
- -0.6$. The emission is absent in the galaxy, so the break in the
H$\beta$ profile at $\log (r/r_{\text{e}}) \approx -0.7$ reflects the
real change of the mean stellar age.

\begin{figure}
\resizebox{\hsize}{!}{\includegraphics{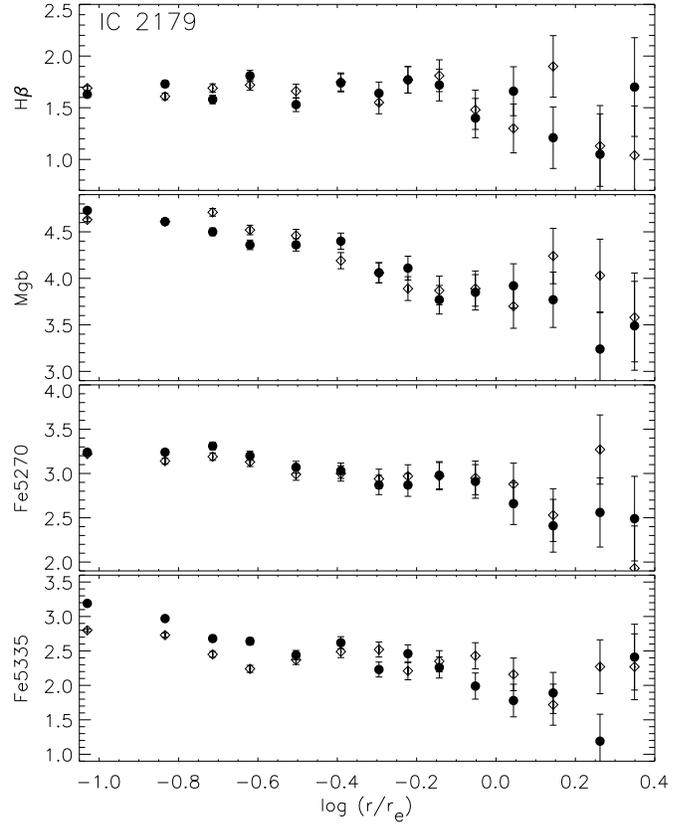}}
\caption{The index profiles of IC\,2179.}
\label{i2179prof}
\end{figure}

{\bf IC\,2179, Fig.~8.} This elliptical galaxy, being the fainter
neighbour of a giant spiral galaxy NGC\,2347, has not been studied in detail
to now.  In particular, the close bright star might preclude any
attempts at surface photometry. We checked the brightness profile
along the slit of the spectrograph and ensured that two halves of the profile
diverge  at $r > 30\arcsec$. However, the surface brightness profile is very
good inside this radius and can be fitted
by a de Vaucouleurs' law \citep{1948AnnAp...11..247V}
 with $r_{\text{e}} \approx 18\arcsec$; if we take only the `fainter'  half
of the profile,  we obtain $r_{\text{e}} \approx 15\arcsec$ in the radius range
between $5\arcsec$ and $56\arcsec$  --  this value is used to parametrise the
profiles in Fig.~8. The metal-line profiles show two breaks -- at $\log
(r/r_{\text{e}}) \approx -0.3 - -0.2$ and at $\log (r/r_{\text{e}})
\approx 0 - +0.1$; the H$\beta$ index profile only reveals the latter
break. The emission lines are absent in the spectrum, so the H$\beta$
index measurements are reliable.

\subsection{The magnesium-to-iron ratios}

\begin{figure*}
\resizebox{15cm}{!}{\includegraphics{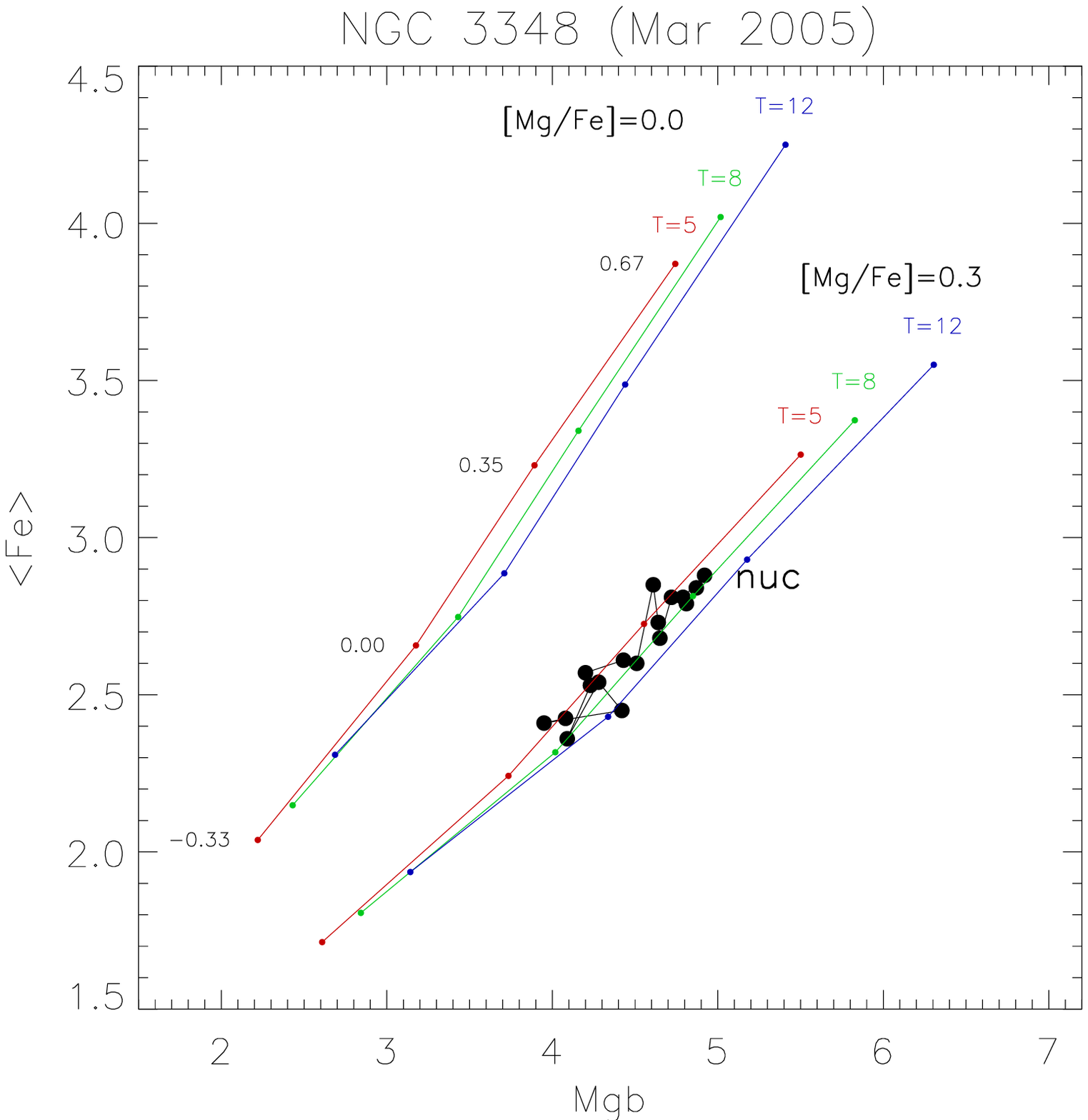}\includegraphics{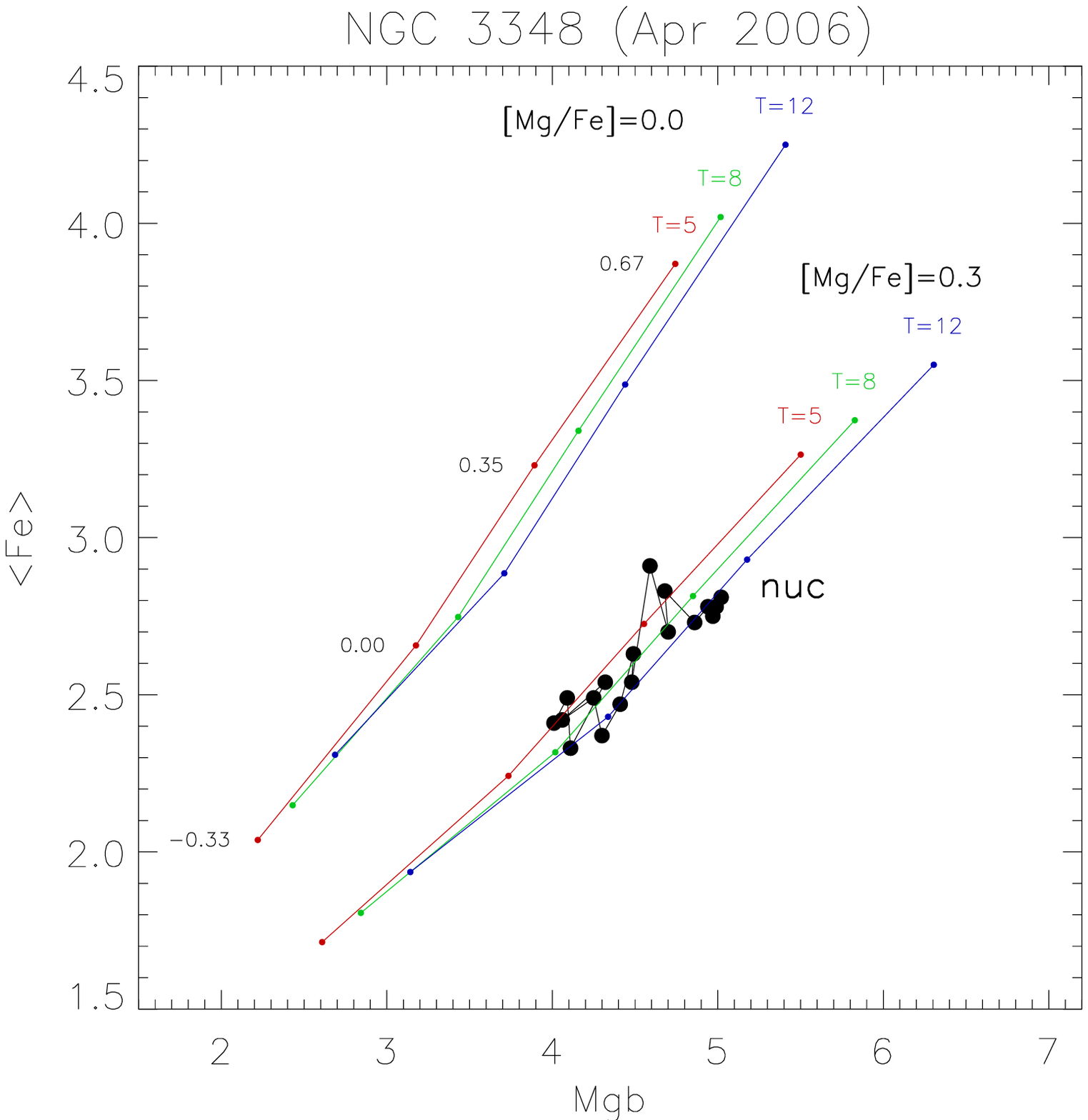}}
\resizebox{15cm}{!}{\includegraphics{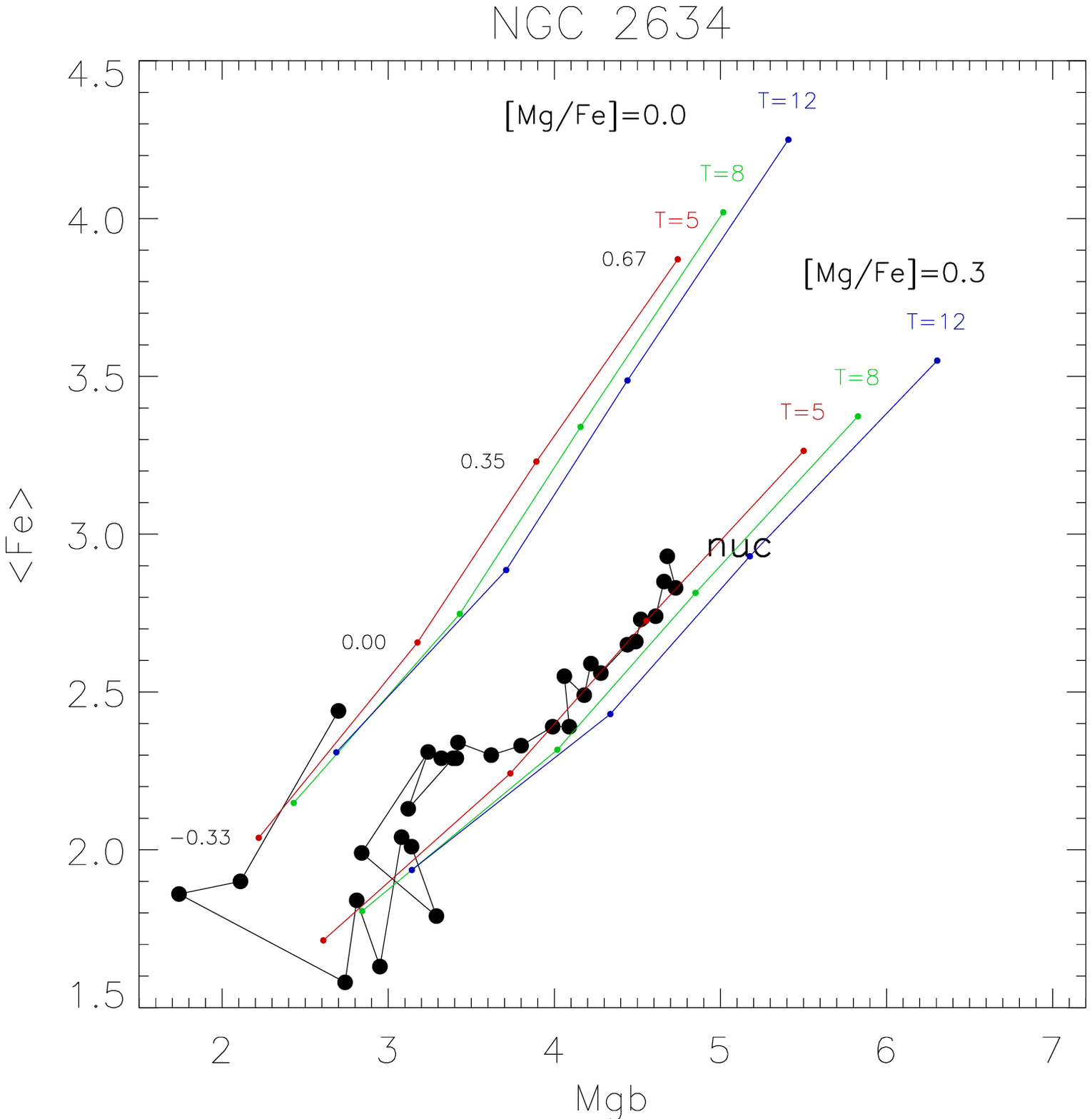}\includegraphics{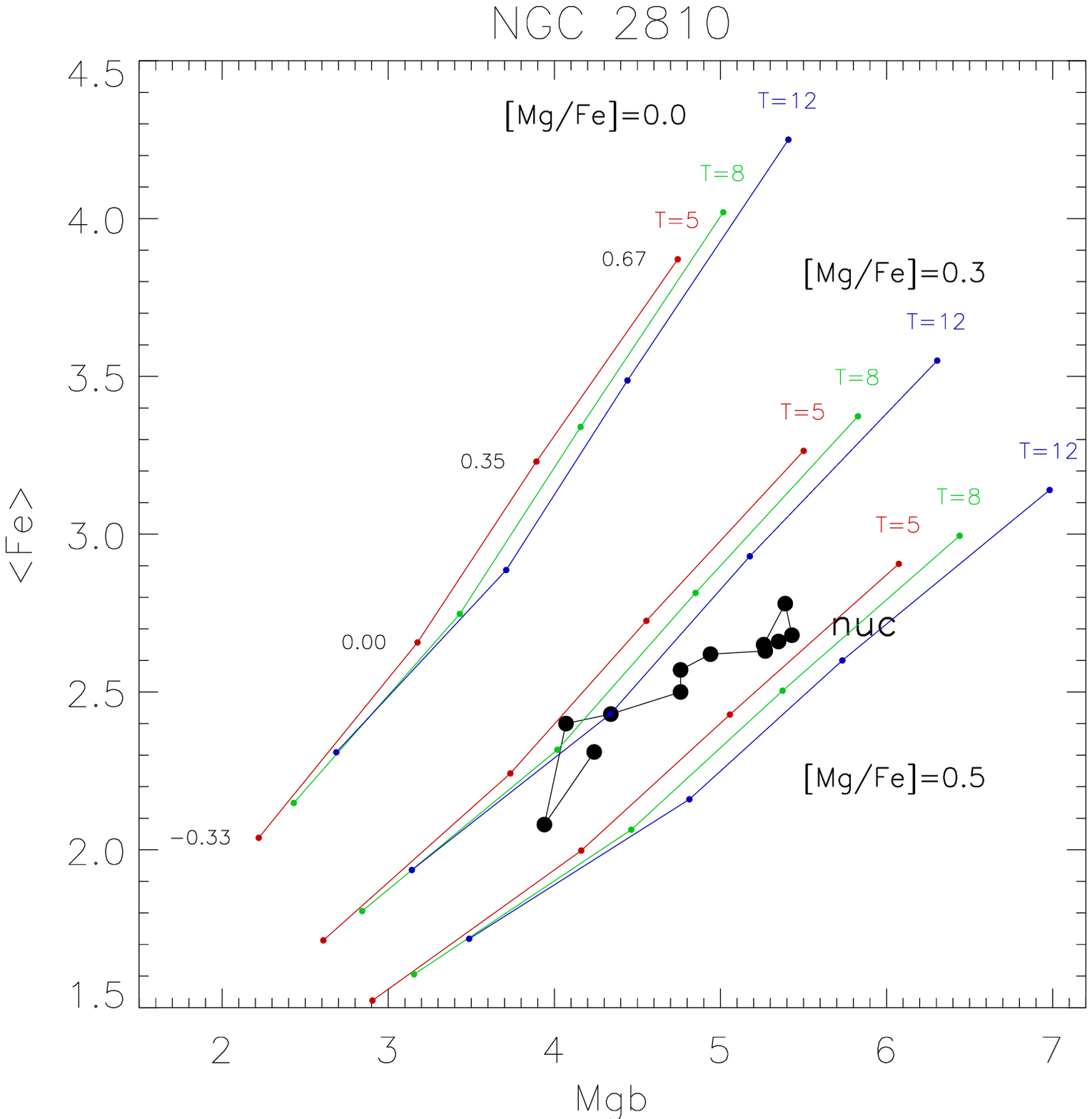}}
\resizebox{15cm}{!}{\includegraphics{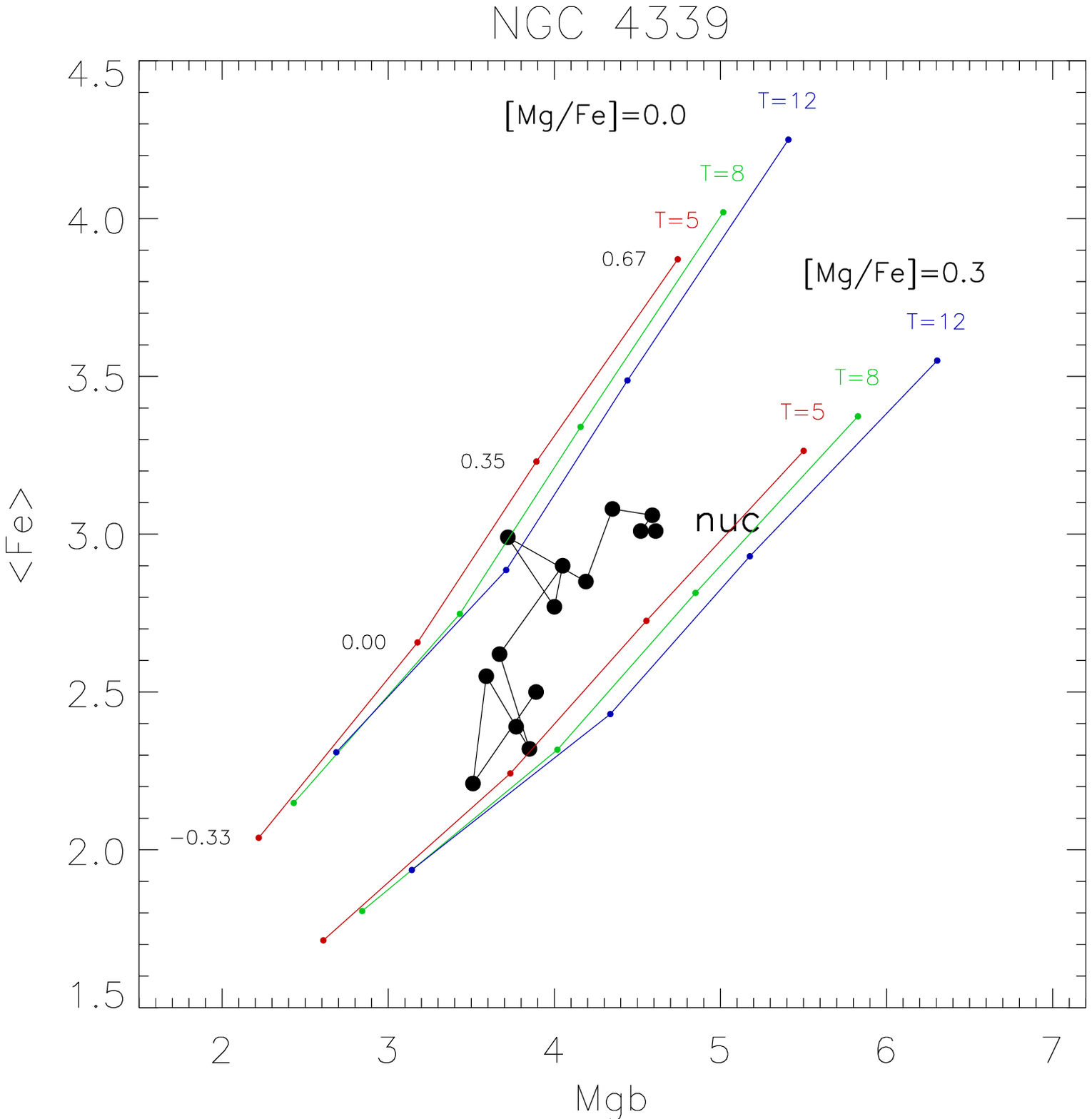}\includegraphics{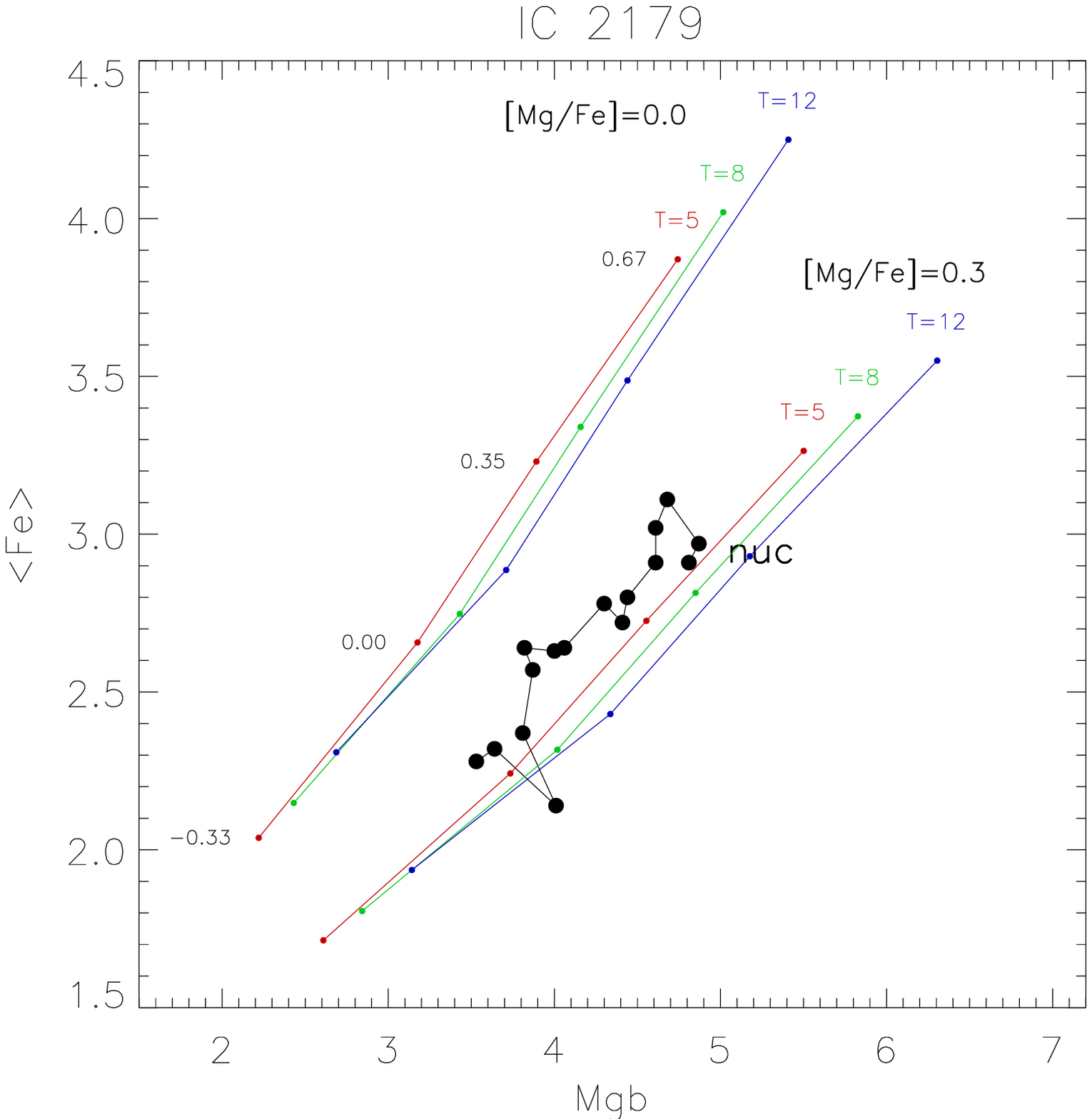}}
\caption{The magnesium-iron index diagrams for all the galaxies under
consideration,  with the Lick indices averaged over both sides of the
slit taken along the radius with the steps as in Figs.~4-8 (circles
connected by solid black lines, the nucleus  marked as `nuc').
The SSP models of \citet{2003MNRAS.339..897T} for [Mg/Fe]=0.0, +0.3,
and +0.5 (the last for NGC\,2810 only) are plotted as a reference
frame; the solid coloured lines represent stellar population models of
equal ages; the metallicities for the models are +0.67, +0.35, 0.0,
--0.33 from the top to bottom of each line.}
\label{mgfe}
\end{figure*}

Figure~9 presents the magnesium-iron diagrams for the index profiles
given in Figs.~4-8. We compare our observational data with the
evolutionary-synthesis models for SSP of \citet{2003MNRAS.339..897T},
which were  calculated for several values of magnesium-to-iron
abundance ratio, namely, the $0.5$, $1$, $2$, and $3$ solar ones
([Mg/Fe]=$-0.3, \, 0.0, \, +0.3$, and $+0.5$). As it is well known
elliptical galaxies have magnesium-overabundant stellar
populations. Our results  for the magnesium-to-iron ratio in the
galaxies under consideration agree with all previous studies of
the stellar population in elliptical galaxies. The mean values
of [Mg/Fe] ratio change from $+0.1$ in NGC\,4339 to $+0.3$ in
NGC\,3348, and our galaxies demonstrate clear dependence of the
magnesium overabundance on the stellar velocity dispersion.  As
already noted by \citet{2000AJ....120..165T}, the more massive
elliptical galaxies have higher magnesium-to-iron ratio. Within
individual galaxies, the [Mg/Fe] ratios remain roughly constant along
the radius as noted earlier by \citet{1992ApJ...398...69W},
\citet{1993MNRAS.262..650D},  and \citet{1995ApJ...448..119F}. The
exception is NGC\,2810 where the centre has almost $\text{[Mg/Fe]} =
+0.5$ -- too high for its stellar velocity dispersion -- and toward the
outer part this ratio falls to its normal value, about $+0.3$.  This
elliptical galaxy has the most prominent emission lines in the
spectrum; and as we see below, the  young age of its central part
is evidence of the secondary star formation burst perhaps
provoked by a minor merger. Less certain change of [Mg/Fe]
from $+0.1$ to $+0.2$ can be noticed along the radius in NGC\,4339.

\subsection{The ages and metallicities along the radius}

In this subsection we present so-called SSP-equivalent ages and
metallicities of the stellar populations determined by fitting the
observed Lick indices H$\beta$, Mg\,b, Fe\,5270, and Fe\,5335 by
one-burst evolutionary synthesis models. The SSP models assume
that all the stars within a stellar system have the same age and metallicity.
In fact, the galaxies may be composite stellar systems formed during
several star formation bursts, with a different chemical composition in
general. We therefore determined some average stellar ages and
metallicities.  The recent simulations by
\citet{2006astro.ph.10343S} have shown that the metallicities
determined in this way are close to luminosity-weighted average
metallicities, while the SSP-equivalent age is always biased toward
the age of the youngest stars within the stellar system and is
somewhere between the luminosity-weighted average age and that of the
youngest stars.

We consider together the H$\beta$ absorption-line index and the metal-line
combined one,
\begin{equation}
\text{[MgFe]}
\equiv
\sqrt{\text{Mg\,b}~\langle \text{Fe} \rangle},
\end{equation}
where
\begin{equation}
\langle \text{Fe} \rangle
\equiv
\frac{\text{Fe}\,5270+\text{Fe}\,5335}{2},
\end{equation}
to disentangle the age and metallicity effects and to determine both
characteristics simultaneously. The model calculations show that, at
the diagram H$\beta$ vs [MgFe],  the line families of equal metallicities
and of equal ages (Figs.~10) have different slopes, so
representing a somewhat curved coordinate frame. By using the model grid
by \citet{2003MNRAS.339..897T} for $\text{[Mg/Fe]} = +0.3$, we applied
a 2D interpolation between the discrete model nodes to calculate the
mean luminosity-weighted stellar ages and metallicities for each
point of the profiles of Figs.~4--8. The H$\beta$ indices in NGC\,2810
and NGC\,3348,  where we see a significant emission contribution,  are
corrected for the emission by using the measurements of the equivalent
width for the emission line [\ion{O}{iii}]\,$\lambda 5007$: the
additive correction is obtained by multiplying
$\text{EW}([\ion{O}{iii}]\lambda 5007)$ by 0.85 for NGC\,2810 (the
individual coefficient estimated from the SCORPIO spectra) and by 0.6
for NGC\,3348 (the statistical coefficient from
\citet{2000AJ....119.1645T}).

To extract main tendencies of age and metallicity variations along the
radius, we go on to average the indices, age,  and metallicity
estimates over wider spatial bins near some key points of the
profiles: $r_{\text{e}}/8$, $r_{\text{e}}/5$, $r_{\text{e}}/2$,
$r_{\text{e}}$, $1.3\,r_{\text{e}}$, and $2\,r_{\text{e}}$. Here we
calculate the errors of the mean indices and those of the mean ages
and metallicities as simple point-to-point scatter divided by the
number of measurements per bin. The results of this procedure are
given in Table~3 and shown in Fig.~10 for each galaxy.

\begin{table*}
\caption{The mean indices, ages, and metallicities along the radius
in the galaxies  of our sample}
\centering
\begin{tabular}{ccccccc}
\hline\hline \\
galaxy &
$\log (r/r_{\text{e}})$ &
H$\beta$ &
Mg\,b &
$\langle\text{Fe}\rangle$ &
[Z/H] &
age [Gyr]
\\ \\ \hline \\
NGC\,2634 & $-\infty $ & $1.55\pm0.02$ & $4.69\pm0.02$ & $2.86\pm0.03$ & $+0.25\pm0.02$ & $11.2\pm0.6$ \\
          & $-0.8$     & $1.51\pm0.03$ & $4.54\pm0.03$ & $2.69\pm0.05$ & $+0.13\pm0.04$ & $12.5\pm0.8$ \\
          & $-0.3$     & $1.61\pm0.01$ & $3.89\pm0.06$ & $2.36\pm0.03$ & $-0.11\pm0.02$ & $13.3\pm0.6$ \\
          & $0.0$      & $1.71\pm0.08$ & $3.25\pm0.08$ & $2.24\pm0.08$ & $-0.17\pm0.03$ &  $9.5\pm0.7$ \\
          & $0.3$      & $1.60\pm0.13$ & $2.78\pm0.06$ & $1.71\pm0.12$ & $-0.5\pm?$     & $11\pm?$ \\
          & $0.45$     & $1.74\pm0.31$ & $2.40\pm0.15$ & $2.17\pm0.16$ & $-0.5\pm?$     & -- \\[2mm]
NGC\,2810 & $-\infty $ & $1.70\pm0.03$ & $5.35\pm0.04$ & $2.71\pm0.02$ & $+0.47\pm0.06$ & $6.5\pm1.3$ \\
          & $-0.7$     & $1.79\pm0.07$ & $5.19\pm0.08$ & $2.63\pm0.06$ & $+0.45\pm0.03$ & $5.3\pm1.0$ \\
          & $-0.3$     & $1.61\pm0.09$ & $4.73\pm0.08$ & $2.50\pm0.04$ & $+0.10\pm0.04$ & $11.2\pm1.8$ \\
          & $0.0$      & $1.82\pm0.05$ & $4.00\pm0.04$ & $2.24\pm0.12$ & $-0.04\pm0.10$ & $8.8\pm1.5$ \\
          & $0.135$    & $1.62\pm0.4$  & $4.24\pm0.2$  & $2.31\pm0.2$  & $+0.1\pm?$     & $12\pm7$ \\[2mm]
NGC\,3348 & $-\infty $ & $1.46\pm0.05$ & $4.92\pm0.06$ & $2.81\pm0.06$ & $+0.30\pm0.02$ & $10.6\pm0.3$ \\
          & $-0.9$     & $1.63\pm0.02$ & $4.78\pm0.05$ & $2.77\pm0.02$ & $+0.29\pm0.02$ & $9.2\pm0.6$ \\
          & $-0.6$     & $1.72\pm0.04$ & $4.57\pm0.03$ & $2.73\pm0.07$ & $+0.26\pm0.04$ & $8.0\pm1.0$ \\
          & $-0.3$     & $1.72\pm0.06$ & $4.21\pm0.04$ & $2.46\pm0.04$ & $+0.04\pm0.01$ & $9.6\pm1.0$ \\
          & $0.0$      & $1.90\pm0.01$ & $4.12\pm0.10$ & $2.44\pm0.02$ & $+0.10\pm0.04$ & $6.6\pm0.4$ \\
          & $+0.12$    & $2.06\pm0.02$ & $4.00\pm0.08$ & $2.61\pm0.2$  & $+0.24\pm0.05$ & $4\pm?$ \\[2mm]
NGC\,4339 & $-\infty $ & $1.61\pm0.03$ & $4.56\pm0.04$ & $3.03\pm0.09$ & $+0.34\pm0.04$ & $9.1\pm1.0$ \\
          & $-0.9$     & $1.81\pm0.03$ & $3.95\pm0.14$ & $2.92\pm0.07$ & $+0.14\pm0.04$ & $7.1\pm0.7$ \\
          & $-0.7$     & $1.70\pm0.10$ & $3.90\pm0.12$ & $2.77\pm0.10$ & $+0.14\pm0.16$ & $7.5\pm2.3$ \\
          & $-0.5$     & $1.59\pm0.07$ & $3.81\pm0.16$ & $2.36\pm0.05$ & $-0.12\pm0.06$ & $11.3\pm2.0$ \\
          & $-0.3$     & $1.70\pm0.06$ & $3.50\pm0.18$ & $2.26\pm0.13$ & $-0.19\pm0.05$ & $11.3\pm0.9$ \\[2mm]
IC\,2179  & $-\infty $ & $1.64\pm0.02$ & $4.78\pm0.02$ & $3.02\pm0.03$ & $+0.44\pm0.02$ & $7.0\pm0.6$ \\
          & $-0.7$     & $1.69\pm0.03$ & $4.55\pm0.05$ & $2.91\pm0.06$ & $+0.33\pm0.03$ & $7.8\pm0.8$ \\
          & $-0.3$     & $1.70\pm0.04$ & $4.12\pm0.07$ & $2.68\pm0.04$ & $+0.10\pm0.04$ & $9.5\pm1.0$ \\
          & $0.0$      & $1.46\pm0.08$ & $3.84\pm0.05$ & $2.47\pm0.10$ & $-0.05\pm0.08$ & $12.2\pm0.4$ \\
          & $+0.3$     & $1.23\pm0.16$ & $3.58\pm0.16$ & $2.30\pm0.2$  & $-0.13\pm?$    & $12\pm?$ \\ \\ \hline\hline
\end{tabular}
\end{table*}

\begin{figure*}
\centering
\resizebox{16cm}{!}{\includegraphics{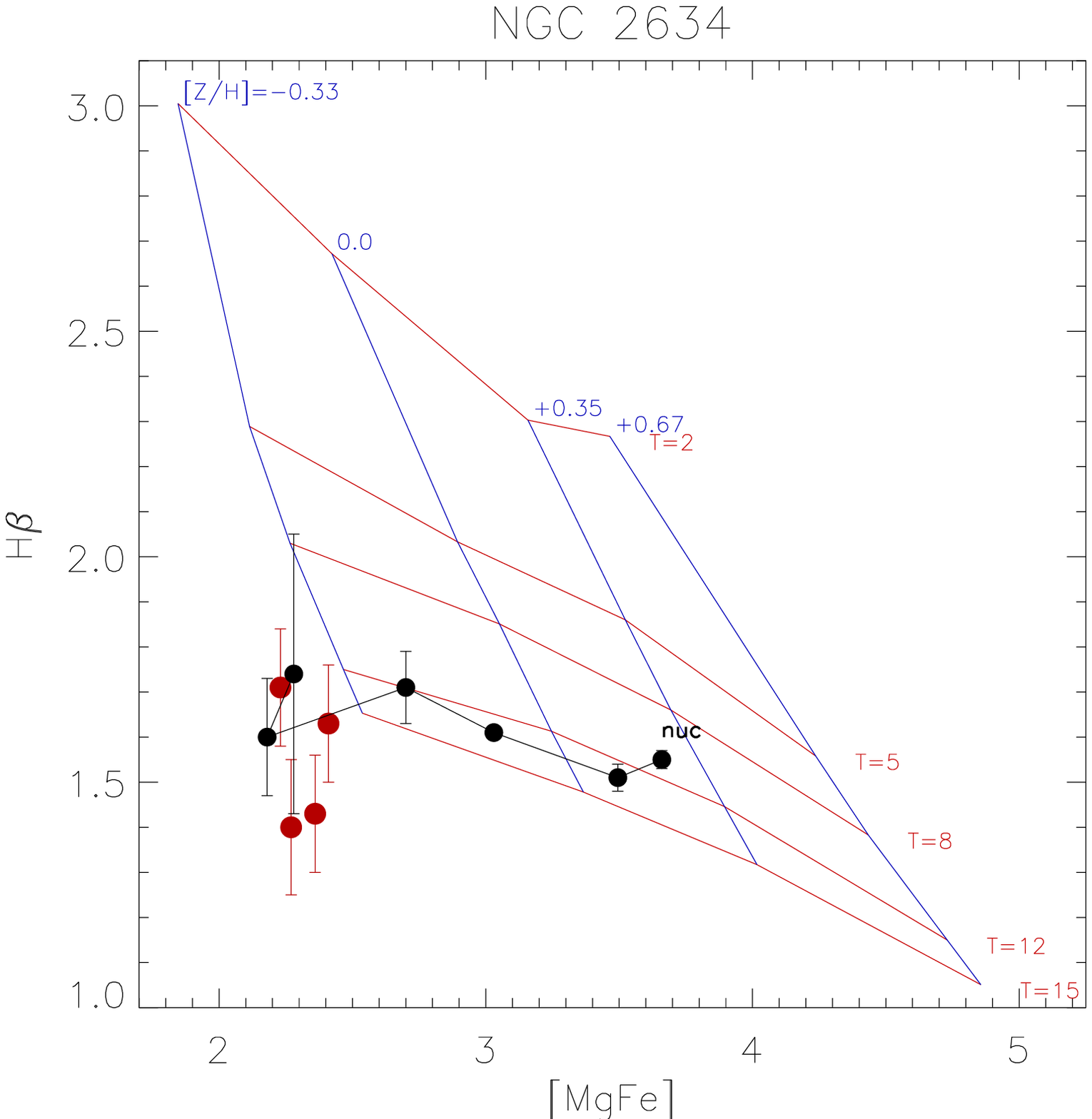}\includegraphics{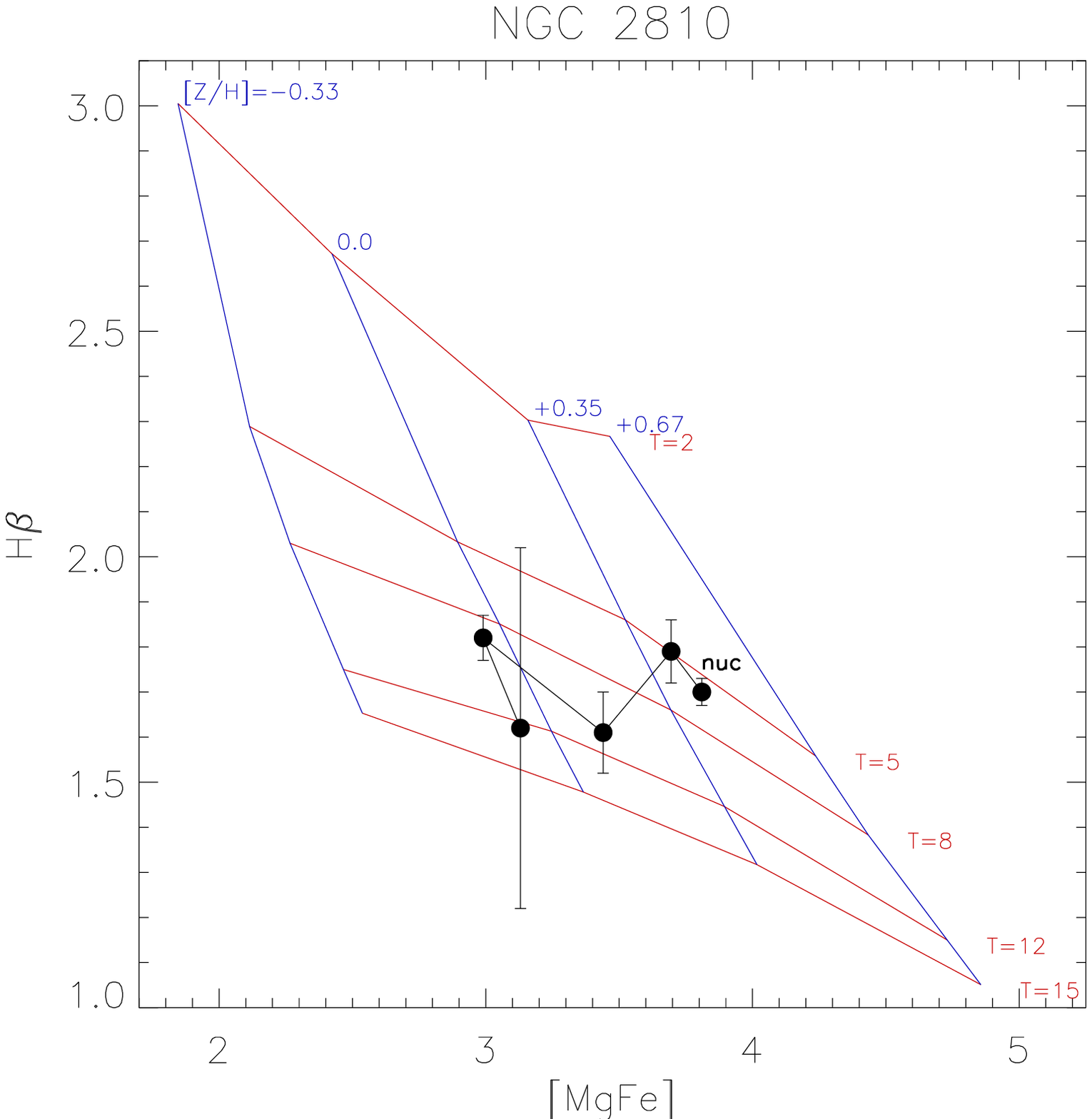}}
\resizebox{16cm}{!}{\includegraphics{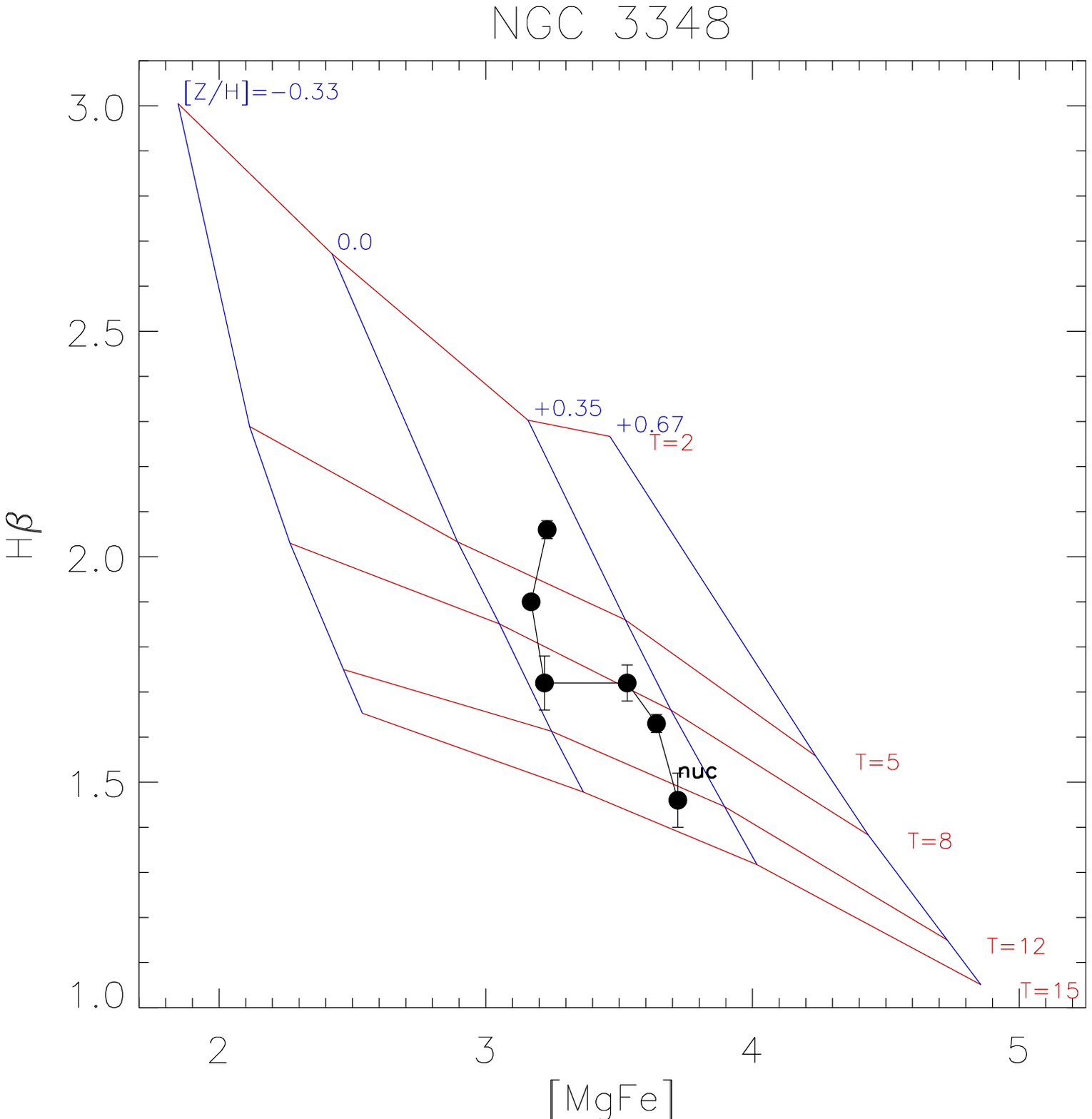}\includegraphics{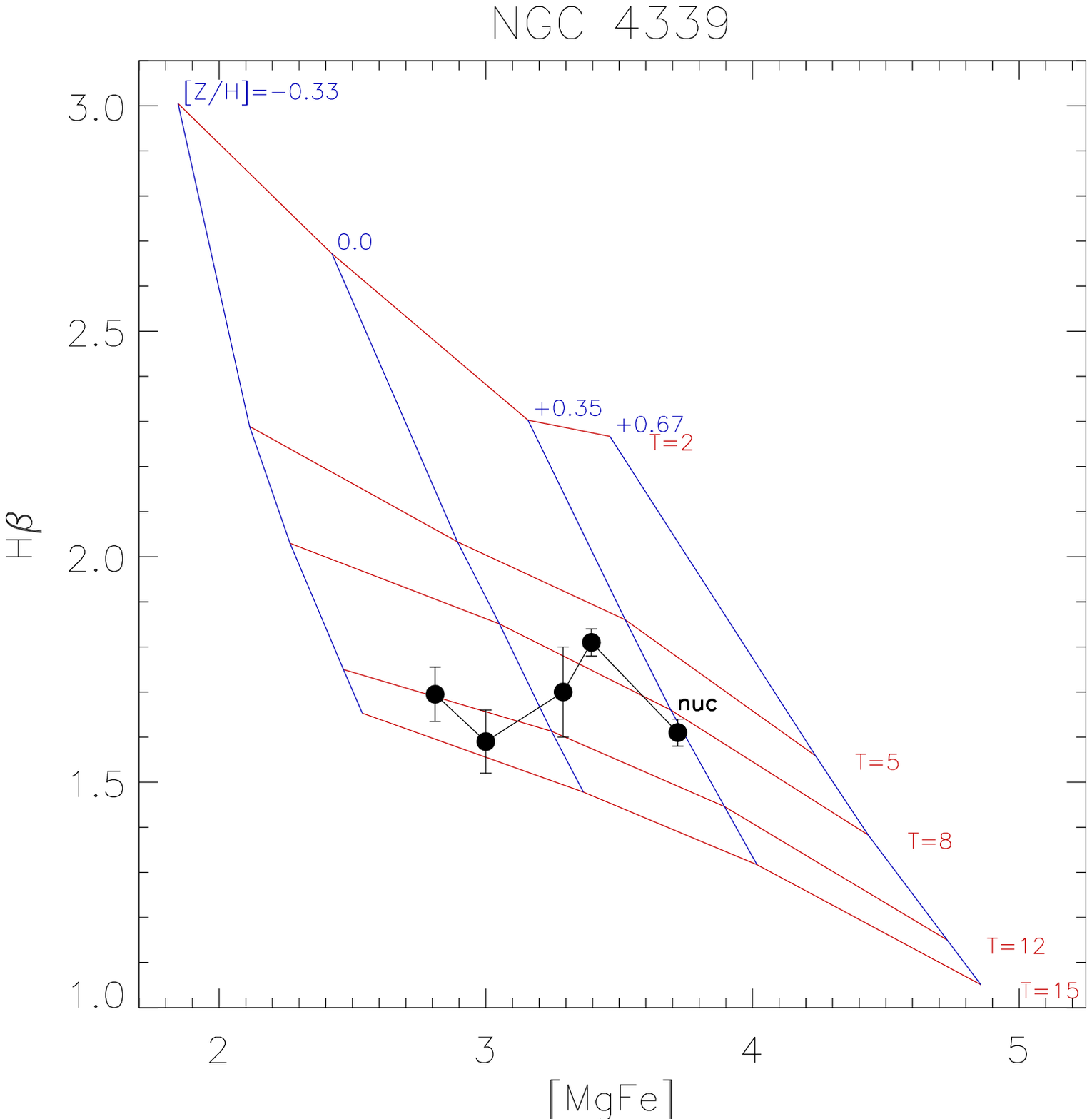}}
\hspace*{3.5cm}
\includegraphics[width=7.5cm]{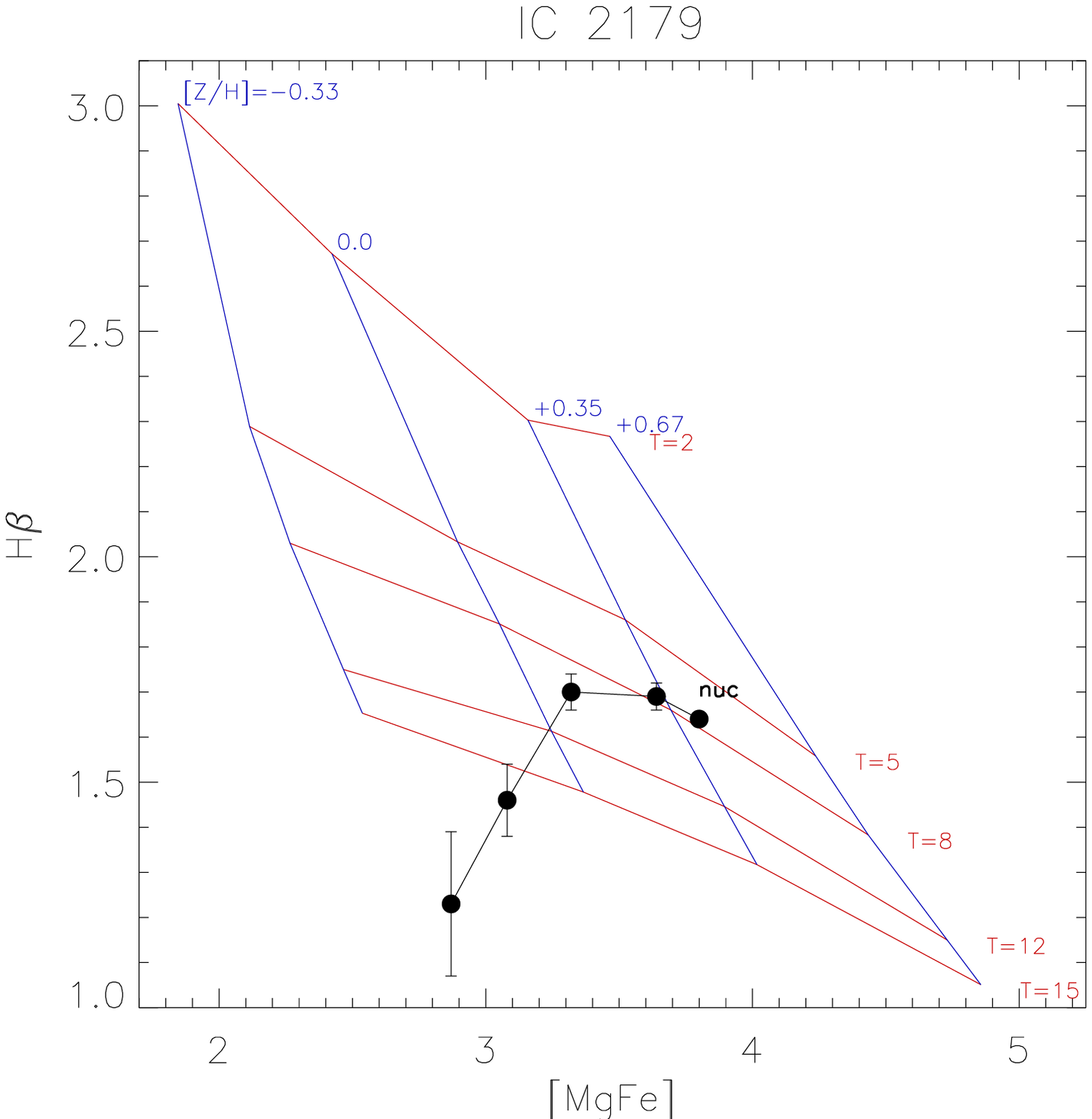}
\caption{The combined metal-line index vs H$\beta$ diagrams for all
the galaxies under consideration with the Lick indices averaged in
wide bins near the key points along the radius -- see the text and the
Table~3 (circles with error bars, the nucleus being marked as `nuc').
The SSP models of \citet{2003MNRAS.339..897T} for [Mg/Fe]$=+0.3$ are
plotted as a reference frame; thin red and blue lines represent
sequences of equal ages and metallicities , correspondingly. At the diagram for
NGC\,2634 several globular clusters of our Galaxy are also plotted as
large red circles.}
\label{ages}
\end{figure*}

Below we discuss  the stellar  metallicity and age variations along the radius
in the individual galaxies.

{\bf NGC\,2634, Fig.~10a.} This only galaxy  in our sample
has no stellar age gradient along the radius. The stellar population
is old everywhere,  10--13~Gyr, and the mean metallicity changes
from roughly $+0.2$ dex in the nucleus to less than $-0.4$ dex at
$r>r_{\text{e}}$. It may seem that the outer measurements of the indices
are outside the parameter range of the models. But this discordance
only reflects the need for more precision in treating low metallicities
by \citet{2003MNRAS.339..897T} (and by any other model sets as
well). For comparison, we have plotted the index measurements for the
Galactic globular clusters \object{NGC\,6624}, \object{NGC\,6838},
\object{NGC\,6356},  and \object{NGC\,6539} with [Fe/H] from --0.35 to
--0.66 from \citet{2004AJ....128.1623B}; they are all near the outer
measurements for NGC\,2634 and outside the model grid of
\citet{2003MNRAS.339..897T}.

{\bf NGC\,2810, Fig.~10b.} The stellar population in the centre of
this field elliptical is not very old, with a mean age of only
5--6 Gyr.  The metallicity of stars in the centre is very high, the
highest in our sample (though the galaxy is not the most massive and
luminous one). Beyond the radius of $3\arcsec$ to $5\arcsec$,  the
stellar population is homogeneously old,  about 10 Gyr, and the
metallicity is approximately solar. As we have also seen, there is a
sharp increase in [Mg/Fe] ratio in the nucleus of NGC\,2810, so by taking
 a considerable amount of ionized gas in this galaxy into account, we
may conclude that a minor merger has taken place some billion years
ago to have provoked a short, intense nuclear star formation burst that
increased the mean metallicity and decreased the mean age of stars.

{\bf NGC\,3348, Fig.~10c.} The galaxy was classified by RSA as a
regular E0, but  it has now shown itself to be very unusual. The radial
variations of the mean stellar age in this galaxy look like a mirror of
NGC\,2810: the nucleus is old, and the stars become younger and younger
along the radius becoming 3--4 Gyr old at $r>r_{\text{e}}$. The metallicity
gradient, though negative at $r=0-0.5\,r_{\text{e}}$,  becomes positive in
the radius range of $0.5\,r_{\text{e}} - 1.3\,r_{\text{e}}$.  Since
NGC\,3348  also shows extended emission lines in the spectrum, we may
suggest that the galaxy has also experienced a minor merger.  But this
time the satellite involved had a smaller amount of gas than that of
NGC\,2810. As a result, no nuclear star formation burst has been
provoked, but the contribution of young and metal-rich stars of the
merged galaxy is clearly seen in the outer part of NGC\,3348. As
evidence in favour of our hypothesis of a minor merger for NGC\,3348,
we demonstrate here the colour map of the galaxy (Fig.~11). We
retrieved the HST/ACS images of NGC\,3348 obtained through two
different filters, F435W and F814W, on January 20, 2003, in the framework
of the programme by W.~E.~Harris on globular cluster systems in
early-type galaxies (Proposal ID 9427).  By dividing one image by
another and by taking 2.5 logarithms of the partial, we obtained  the
colour map, close to $B-I$,  that is presented in Fig.~11. One can see
a lot of thin red filaments irregularly distributed over the whole
extension of the galaxy. Evidently, this `prototype of E0', as
\citet{1994cag..book.....S} characterized NGC\,3348 in their
Atlas, contains in fact a noticeable amount of diffuse dust and ionized
gas coupled with this dust , which may be accreted rather recently.

\begin{figure}
\resizebox{\hsize}{!}{\includegraphics{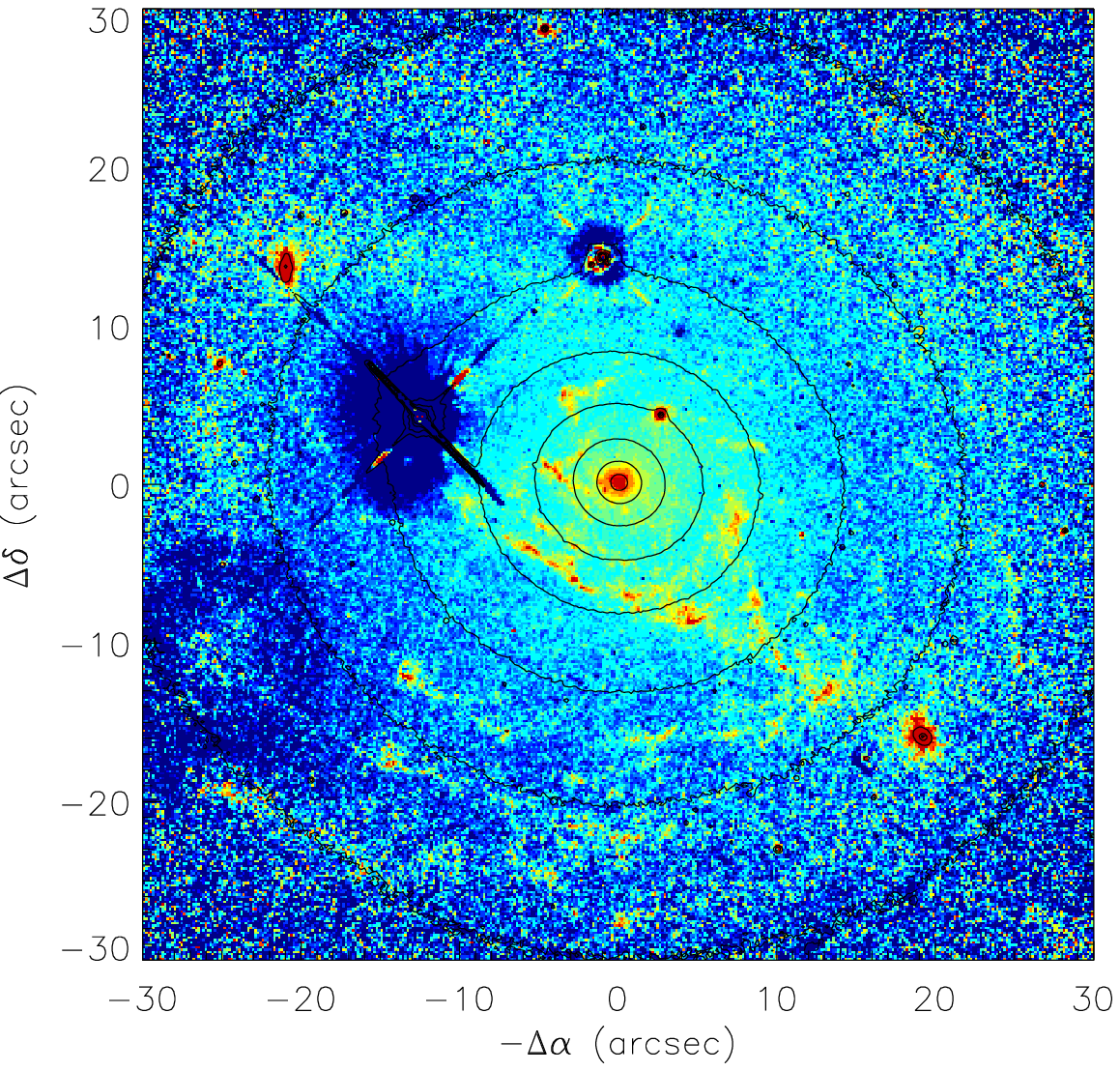}} \caption{The colour
F435W-F814W (HST/ACS) map of NGC\,3348.} \label{colour3348}
\end{figure}

{\bf NGC\,4339, Fig.~10d.} Though less prominent than NGC\,2810,
NGC\,4339 has also shown some signs of the circumnuclear young
metal-rich stellar disk that  may be a product of the secondary
nuclear star formation burst. However, the gas is absent in this
galaxy.

{\bf IC\,2179, Fig.~10e.} In IC\,2179 the nucleus is also younger than
the outer part; but unless NGC\,2810 and NGC\,4339, IC\,2179 does not
seem to have a young circumnuclear stellar disk producing a `step' in
age radial variations in the former galaxies. Instead we observe the
gradual smooth age increase along the radius, from 8 Gyr to $\sim 15$
Gyr toward $2\,r_{\text{e}}$.  The metallicity decrease along the
radius is also very smooth, since IC\,2179 is the only galaxy in our sample
where we do not see flattening of the metallicity gradient at
$r>0.5\,r_{\text{e}}$.

\section{Discussion}

The main point of our discussion concerns the metallicity gradients in
elliptical galaxies,  which are one of the signatures of their origin.
As  already mentioned, the monolithic dissipative collapse of a
protogalactic cloud produces steep stellar metallicity
gradients along the radius: from $-0.5$ \citep{1984ApJ...286..403C} to
$-1.0$ dex per radius dex \citep{2004MNRAS.347..740K}. As the mean
metallicity gradients obtained to date under the assumption of the
constant stellar age along the radius from the colour and metal-line
index gradients was about $-0.2$ dex per radius dex, the opinion has
formed that a major merger is typical of  a luminous elliptical galaxy
to flatten its metallicity gradient.  Refining this zero-order
approach, \citet{2004MNRAS.347..740K}, \citet{2005MNRAS.361L...6F}, and
\citet{2005ApJ...632L..61O} all note a wide spread of metallicity
gradients just for luminous (most massive) elliptical galaxies,  thereby
suggesting that there is a variety of evolutionary histories for
elliptical galaxies, from quasi-monolithic collapse (or early gas-rich
multiple mergers) to `dry' major mergers.

In our work, we have disentangled the age and metallicity effects and
measured both characteristics of the stellar populations as they
change along the radius. First of all, in most galaxies the stellar
age \emph{is not} constant along the radius: its behaviour is quite
various and does not depend on the galaxy's  luminosity or
environment. Secondly, when we have carefully analysed  the
metallicity profiles over the large radial base, to at least
$1.3\,r_{\text{e}}$, we find that the metallicity variations
cannot be described by a single power law over the full radial
extension in  most galaxies. The break of the slope is evident at
$\sim 0.5\,r_{\text{e}}$ the outer parts showing the flatter
metallicity gradients. We calculated the metallicity gradients
for our 5 galaxies separately within $r<0.5\,r_{\text{e}}$ and in the
radius range of $0.5\,r_{\text{e}} - 1\,r_{\text{e}}$, and  the results are
given in Table~4 with the galaxies ranked according to their stellar
velocity dispersion (mass).

\begin{table}
\caption{Metallicity gradients in our  galaxies }
\centering
\begin{tabular}{ccc}
\hline\hline \\
galaxy & d[Z/H]$/$d$\log r$ & d[Z/H]$/$d$\log r$ \\
       & ($r<0.5\,r_{\text{e}}$) &  ($0.5\,r_{\text{e}}<r<r_{\text{e}}$) \\ \hline \\
NGC\,3348 & $-0.42\pm0.05$ & $+0.20\pm0.17$ \\
NGC\,2810 & $-0.88\pm0.17$ & $-0.47\pm0.47$ \\
IC\,2179  & $-0.58\pm0.18$ & $-0.50\pm0.40$ \\
NGC\,2634 & $-0.52\pm0.16$ & $-0.20\pm0.17$ \\
NGC\,4339 & $-0.82\pm0.52$ & -- \\ \\
\hline\hline
\end{tabular}
\end{table}

From Table~4 one can see that the metallicity gradients in the inner
parts of the galaxies are rather steep, steeper than the $-0.4$ dex per
radius dex, and are consistent with the model of monolithic
dissipative collapse. In the outer parts,  the metallicity gradients are
consistent with being zero; also taking the very inhomogeneous mean
age distributions at $r>0.5 \,r_{\text{e}}$ into account, we may
suggest that the outer parts of the elliptical galaxies have been
transformed in a secular manner through minor mergers, not by a
\emph{major} merger. This picture differs strongly from the
conventional scenario of elliptical galaxy formation.

An interesting question is  the possible connection between metallicity and
age variations. Earlier, \citet{1999PASP...111..919H} noted an
anticorrelation between the metallicity and age gradients found
over the galaxy sample typically measured to $0.5-1 \,r_{\text{e}}$.
The slope of this relation appeared to be close to $-\frac{3}{2}$,
which was a ratio of the sensitivities of the majority of metal-line indices
to the age and metallicity variations,
$\Delta \mbox{Age} / \Delta \mbox{Z}$ \citep{1994ApJS...95..107W}.
\citet{1999PASP...111..919H} propose that (i) all the differences between
the measured metallicity and age gradients in elliptical galaxies are due
to random and systematic errors in index evaluation, producing the
correlated scatter of the measured metallicity and age gradients
and (ii) that, in fact,
all the elliptical galaxies have to have exactly  d[Z/H]$/$d$\log r =-0.25$
and  d$\log T/$d$\log r =+0.1$. The accuracy of our measurements
of the Lick indices along the radius of elliptical galaxies is higher
than earlier results. We are therefore confident  that
the difference between the inner and outer metallicity
gradients, say, in NGC\,3348 or NGC\,2634, is  statistically significant.
But indeed, in our sample too, the metallicity and age gradients show a certain
anticorrelation (if we exclude the very outer part of NGC\,2810),  with
a slope of $-1.20\pm 0.32$. This is  coincident with $-\frac{3}{2}$
within the error. Although this correlation may be real, its physical
origin remains uncertain.  A possible qualitative explanation is  that
additional star formation would simultaneously produce
additional metals, while lowering the mean stellar age.

Another important point concerning the observational signatures  of the
dominant mechanism for  elliptical galaxy formation is correlation
between the dynamical parameters, mass or escape velocity, and the
mean abundance characteristics of the stellar population. Initially,
the well-known correlation between luminosity and the colour of elliptical
galaxies was treated as mass-metallicity relation
\citep{1973ApJ...179..731F} similar to what is known for other types of
galaxies \citep[e.g.][]{2004A&A...425..849P}.  But central-aperture
spectral investigations of complete samples of nearby elliptical
galaxies have given somewhat unexpected results: when disentangling
the age-metallicity degeneracy and simultaneously measuring age,
metallicity, and the [$\alpha$/Fe] ratio, \citet{2000AJ....120..165T} and
recently \citet{2005AJ....130.2065H} have noted a strong correlation
between the [$\alpha$/Fe] ratio and the stellar velocity dispersion,
but the metallicity--$\sigma _*$ correlation appears to be
weak. This fact provokes a revision of the formation models for elliptical
galaxies. For example, \citet{2004MNRAS.347..968P} introduce an
`inverse wind' or a direct relation between the stellar velocity
dispersion and star formation efficiency in every location within a
proto-elliptical galaxy.  High star-formation efficiency in the
densest regions means the shortest epoch of the main star formation,
hence the highest [$\alpha$/Fe] ratio that is consistent with the
observational data; however, the physical base of this relation remains
unclear.

\begin{figure}
\resizebox{\hsize}{!}{\includegraphics{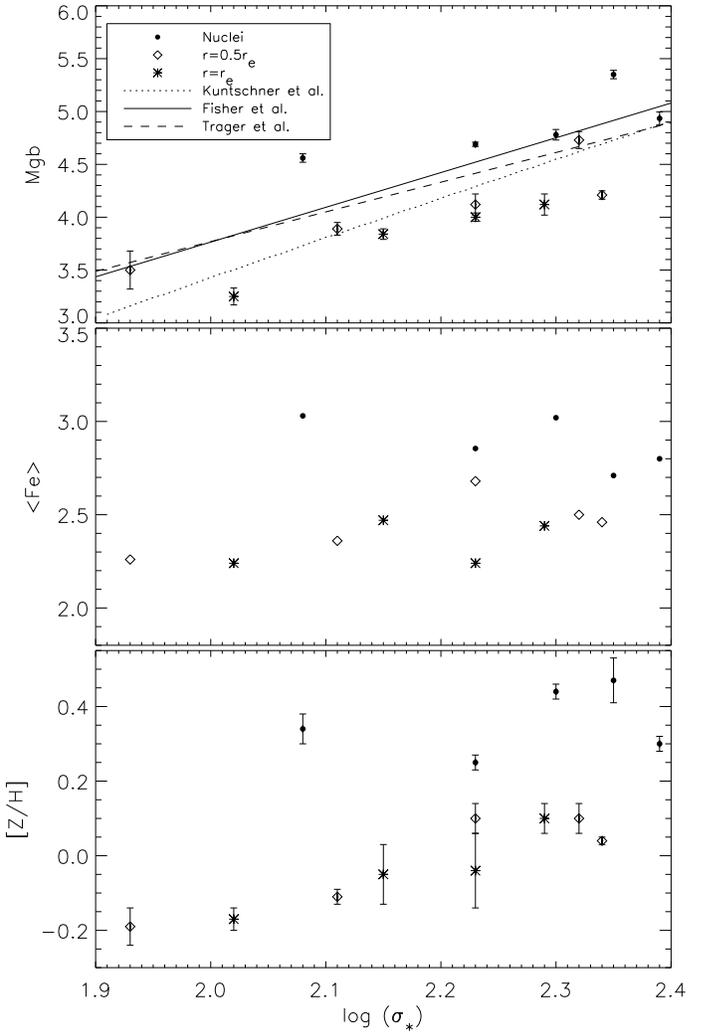}}
\caption{The correlation between the Mg\,b and $\langle \text{Fe}
\rangle$ indices and metallicity and the local stellar
velocity dispersion in three positions along the
radii in galaxies: in the nuclei, at $r=0.5\,r_{\text{e}}$, and at
$r=r_{\text{e}}$.}
\label{correl}
\end{figure}

We have plotted our results concerning the relation between the
metal-line indices and the mean stellar metallicity with the stellar
velocity dispersion in Fig.~12;  we have
plotted not only the nuclei, but also the characteristics at
$r=0.5\,r_{\text{e}}$ and at $r=r_{\text{e}}$. In the top plot, where
we compare the Mg\,b index and the stellar velocity dispersion,  we
also indicate the mean relations found earlier by
\citet{1995ApJ...448..119F}, \citet{2000AJ....120..165T}, and
\citet{2001MNRAS.323..615K}. The high accuracy of our data allows
us to separate  dependencies for the nuclei and for the off-nuclear
regions.  They go in parallel, with a significant vertical shift.
Meanwhile, the relations published so far go much steeper than our
observations and `try' to involve all the points  that  are in fact
physically separated. The clearest physical picture is shown by
the bottom plot, [$Z/$H] vs $\log \sigma$. The points for
$r=0.5\,r_{\text{e}}$ and $r=r_{\text{e}}$ are mixed; they represent
good correlation, the rms scatter around the linear regression,
$[Z/\text{H}]=-1.54+0.69\log \sigma$, is only 0.04 dex.  In
contrast, the points for the nuclei are certainly above this relation,
and for the nuclei we do not see any correlation between the (nuclear)
metallicity and the central stellar velocity dispersion.

The separated position of the nuclear points in Fig.~12 forces us to
suggest that the \emph{nuclear} star formation in elliptical galaxies is
never stopped by galactic wind, but instead proceeds quickly and efficiently
toward the full exhaustion of the gas.  The duration of star formation
over the whole galactic body may be limited by the onset of a galactic
wind due to the feedback from hot stars and supernovae and is governed by
gravitation depending on the depth of the potential well.  In any
case, our results imply that all the physical correlations
must be searched for only among the integrated characteristics of
elliptical galaxies. Their stellar nuclei have their own evolution and
are not representative of the galaxies as a whole.

\section{Conclusions}

By studying the Lick index gradients in 5 elliptical galaxies with
moderate luminosity and located in different environments, we have
carefully disentangled age-metallicity effects and  measured
radial variations of the mean luminosity-weighted stellar ages and
metallicities up to $0.5\,r_{\text{e}}$ in one galaxy, up to
$1.3\,r_{\text{e}}$ in two galaxies and up to $2\,r_{\text{e}}$ in
another two galaxies.  We have found that the mean stellar age is
constant along the radius only in one galaxy out of the 5. The other 4
galaxies demonstrate quite different mean stellar age behaviour: the
outer parts are older than the centres in 3 cases and younger  in
one case.  Among the former, we suspect the influence of the
compact, rather young circumnuclear disks in two galaxies.  The
metallicity gradients cannot be approximated by a single power law
over the full extension of the radius in 4 galaxies out of 5. The inner
metallicity gradients within $0.5\,r_{\text{e}}$ are all rather steep,
steeper than --0.4 metallicity dex per radius dex, and are
inconsistent with the origin of the elliptical galaxies by a major
merger. The outer parts may represent mostly the mix of stellar
populations provided by minor mergers.

\begin{acknowledgements}
During the data analysis we have used the Lyon-Meudon Extragalactic
Database (LEDA) supplied by the LEDA team at the CRAL-Observatoire de
Lyon (France) and the NASA/IPAC Extragalactic Database (NED) which is
operated by the Jet Propulsion Laboratory, California Institute of
Technology, under contract with the National Aeronautics and Space
Administration.  The research is partly based on observations made
with the NASA/ESA Hubble Space Telescope, obtained from the data
archive at the Space Telescope Science Institute, which is operated by
the Association of Universities for Research in Astronomy, Inc., under
NASA contract NAS 5-26555.  The spectral study of stellar populations
in early-type group and cluster galaxies is supported by the Russian
Foundation for Basic Researches (05-02-19805-MFa).
\end{acknowledgements}

\end{document}